\documentclass[journal,twoside]{IEEEtran}
\usepackage{multirow}
\usepackage{booktabs}
\usepackage{amsfonts}
\usepackage{amsmath,cases}
\usepackage{amssymb}
\usepackage{graphicx}
\usepackage{cite}
\usepackage{epsfig}
\usepackage{url}
\usepackage{algorithm}
\usepackage{algorithmic}
\usepackage{epstopdf}
\usepackage{color}
\usepackage[tight]{subfigure}
\usepackage{balance}
\DeclareMathAlphabet\mathbfcal{OMS}{cmsy}{b}{n}

\newcommand{\st}{{\mathrm{s.t.}}}

\newcommand{\ds}{\displaystyle}

\newcommand{\clL}{{\cal L}}

\newcommand{\pk}{p^{(\kappa)}}
\newcommand{\alphak}{\alpha^{(\kappa)}}
\newcommand{\betak}{\beta^{(\kappa)}}
\newcommand{\chik}{\chi^{(\kappa)}}
\newcommand{\deltak}{\delta^{(\kappa)}}
\newcommand{\gammak}{\gamma^{(\kappa)}}
\newcommand{\fk}{f^{(\kappa)}}

\ifCLASSOPTIONpeerreview

\fi

\begin{document}
\setcounter{page}{1}

\title{Full-Duplex MIMO-OFDM Communication with Self-Energy Recycling}
\author{Ali A. Nasir,  Hoang D. Tuan, Trung Q. Duong, and H. Vincent Poor
\thanks{A.~A.~Nasir is with the Department of Electrical Engineering, King Fahd University of Petroleum and Minerals (KFUPM), Dhahran, Saudi Arabia (email: anasir@kfupm.edu.sa).}
\thanks{H.~D.~Tuan is with the School of Electrical and Data Engineering, University of Technology Sydney, Broadway, NSW 2007, Australia (email: Tuan.Hoang@uts.edu.au).}
\thanks{T.~Q.~Duong is with Queen's University Belfast, Belfast BT7 1NN, UK  (email: trung.q.duong@qub.ac.uk)}%
\thanks{H.~V. Poor is with the Department of Electrical Engineering, Princeton University, Princeton, NJ 08544, USA (email: poor@princeton.edu).}
}

\maketitle

\vspace*{-1.0cm}

\begin{abstract}
This paper focuses on energy recycling in full-duplex (FD) relaying multiple-input-multiple-output orthogonal frequency division multiplexing (OFDM) communication. The loop self-interference (SI) due to full-duplexing is seen as an opportunity for the energy-constrained relay node to replenish its energy requirement through wireless power transfer. In forwarding the source information to the destination, the FD relay can simultaneously harvest energy from the source wireless transmission and also through energy recycling from its own transmission. The objective is to maximize the overall spectral efficiency by designing the optimal power allocation over OFDM sub-carriers and transmit antennas. Due to a large number of sub-carriers, this design problem poses a large-scale nonconvex optimization problem involving a few thousand variables of power allocation, which is very computationally challenging. A new path-following algorithm is proposed, which converges to an optimal solution. This algorithm is very efficient since it is based on \textit{closed-form} calculations. Numerical results for a practical simulation setting show promising results by achieving high spectral efficiency.

\end{abstract}

\begin{IEEEkeywords}
Full-duplex relaying, MIMO-OFDM, energy recycle, spectral efficiency, large-scale nonconvex optimization, power allocation.
\end{IEEEkeywords}

\section{Introduction}

\subsection{Motivation and Literature Survey}\label{sec:mot}

Full-duplexing is an advanced communication technology, which enables simultaneous signal transmission and reception at the same  communication node at the same time and over the same frequency band \cite{Zhang-15-May-A}. The major issue in full-duplex (FD) communication is the loop self-interference (SI) due to the co-location of transmit and receive antennas. For instance, with the current state-of-the-art  technology of signal isolation and rejection, it is still not practical to mitigate the FD SI to a level worthy for simultaneous high uplink and downlink throughputs \cite{Zhang-16-Jul-A}. This is  because a hardware non-linearity, e.g., phase noise, may significantly limit the combined analog and digital SI cancellation, thus resulting in a high residual SI. However, such high-powered interference of the signal transmission to the signal reception can open an opportunity for the simultaneous information transmission and energy-harvesting \cite{Lu-14-A}. Indeed, unlike the information throughput, which is dependent on the signal-to-interference-plus-noise-ratio (SINR) and considers the interference as a foe, the energy-harvesting is dependent on the received signal power only and hence, it considers interference as an energy source. 
Therefore, a scenario of transmitting information while harvesting energy through self-energy recycling or other means is really attractive in exploring the potential of full-duplexing \cite{Nasir-13-Jul-A,ZZ15,Zhang-17-Oct-A}.
A brief literature survey discussing research issues, challenges, and opportunities of wireless power transfer-aided FD relay systems is provided by \cite{Wei-18-A}. Another survey discussing the opportunities and challenges of multi-antenna beamforming techniques in FD and self energy recycling is provided by \cite{Hwang-17-Oct-A}. The communication via FD relaying can occur in two phases or it may happen in a single phase via power splitting (PS).

A truly FD communication with wireless energy harvesting (EH) is achieved by employing PS at the relay, which
splits the received signal for EH and information decoding (ID) and simultaneously forwards the information signal to the destination \cite{Wang-17-Jun-A,Zhao-17-Oct-A,Wang-16-Oct-A,Chen-17-A,Wang-17-A,Liu-16-Dec-P,Orikumhi-17-Dec-A}. Different communication setups, e.g., single-input-single-output (SISO) communication \cite{Wang-17-Jun-A}, SISO communication with multiple relays and relay selection \cite{Chen-17-A}, multiple-input-single-output (MISO) communication \cite{Zhao-17-Oct-A,Wang-16-Oct-A}, two-way relaying with multiple-antenna relay and FD source nodes \cite{Wang-17-A},multiple-input multiple-output (MIMO) communication \cite{Liu-16-Dec-P,Orikumhi-17-Dec-A}, have been considered. However, the PS approach is complicated and inefficient for practical implementation due to the requirement of variable power-splitter design \cite{Nasir-16-TCOM-A}. In addition, as mentioned above, it is still challenging with the current state-of-the-art electronics to attenuate SI to such a degree which could enable the PS-based FD system to provide promising spectral efficiency. A practical option is to go with two-phase communication where in the first phase, the source transmits the information signal to the relay and during the second phase, the source transmits the energy signal to the relay and the FD relay, simultaneously, forwards the source information to the destination. Thus, the relay, being an energy constrained node, can replenish its energy by means of wireless energy signal from the source and also through energy recycling from its own transmission \cite{ZZ15}. Using such protocol, different  communication setups, e.g., SISO communication from the source-to-relay and MISO communication from the relay-to-destination \cite{Kim-16-Jan-P}, single-input-multiple-output (SIMO) communicaiton from the source-to-relay and MISO communicaiton from the relay-to-destination \cite{Mohammadi-Apr-16-A,Yadav-Jul-18-A}, SISO communication over both links \cite{Liu-16-Nov-A,Hu-16-A,Chen-18-A,Su-17-A}, MISO communication over both links \cite{ZZ15,Wu-17-Apr-A}, SISO communication with multiple relays and relay selection \cite{Le-18-Feb-A}, MIMO communication from the source-to-relay and MISO link from the relay-to-destination \cite{Demir-18-Feb-A}, and MIMO communication over both links \cite{Zhang-17-Oct-A}, have been considered.

{One of the main issues of MIMO communication is its channel frequency selectivity
due to multipath propagation, which is overcame by  the orthogonal frequency division multiplexing (OFDM) technology.
By transforming the frequency selective channel into parallel frequency flat sub-channels,  
MIMO-OFDM provides the dominant air interface for broadband wireless communications \cite{Gold05,Cho-B-10}. } 
 The communication range can be further extended by deploying cooperative relaying based communication to assist the transmission between the two distant ends (see e.g. \cite{Hammerstrom-07-Aug-A,NgSchoberTVT-10} for half-duplex (HD) relaying and \cite{Ng-12-May-A,TNT17} for  FD relaying).
{Wireless information and power transfer in  HD OFDM relay networks is considered in \cite{Ng-13-Dec-A,Zhou-14-Apr-A,Xiong-Aug-15-A}.} 
In a nutshell, the focus of this article is on an MIMO-OFDM system with FD relaying, in which the cooperative relay forwards the source information to the destination and in the meanwhile, replenishes or harvests energy not only by wireless signals from the source but also through energy recycling from its own transmission. 

\subsection{Research Gap and Contribution:}

In this paper, {we consider the practical two-phase communication, as described just above, 
which implements MIMO-OFDM to handle the frequency selectivity of the both MIMO channels from the source to the
relay and from the relay to the destination.}
The relevant research in the existing literature, but assuming the both channels flat, has been carried out in \cite{Zhang-17-Oct-A}. Specifically, it considered the design of robust non-linear transceivers (source
precoding, MIMO relaying, and receiver matrices) to minimize the mean-squared error at the destination end
in the presence of channel uncertainty. Alternating optimization, which optimizes one of these three matrices with
other two held fixed, is used to address the posed optimization problem. The alternating optimization in the source precoding matrix is still nonconvex, for which the d.c. (difference of two convex functions/sets) iteration (DCI)
\cite{KTN12} is revoked. It should be noted that the solution founded by alternating optimization does not necessarily
to satisfy a necessary optimality condition. Such an optimization problem as posed in \cite{Zhang-17-Oct-A} should be much more efficiently addressed by
the matrix optimization techniques in \cite{RTKN14,Taetal17}, which simultaneously optimize all the matrix variables in each iteration, avoiding the alternating optimization.

It is worth mentioning that under MIMO-OFDM, the design of  precoding and relaying matrices is simplified
to that of power allocation over multiple subcarriers and transmit antennas. Our objective is to maximize the spectral efficiency in terms of sum rate. 
This optimization problem of power allocation is not only nonconvex but
is large-scale due to a large number of subcarriers (up to thousands), making all the existing nonconvex solvers useless. {For example, the aforementioned DCI \cite{KTN12} can hardly solve the problem (demanding huge computational time) only for small-scale systems with a few number of sub-carriers.}
The main contribution in this paper is to propose a new path-following algorithm, which is practical for large-scale nonconvex
optimization of thousands of variables (sub-carriers). We develop several innovative approximations for both concave and nonconcave functions, and for both convex and nonconvex sets which enable the iterative optimization problem to admit the optimal solution in a closed-form. Our approach guarantees a computational solution, around in one-tenth of a second, even in the presence of thousands of subcarriers, due to quick convergence and closed-form calculations over each iterations. {Our detailed computational complexity analysis clearly shows huge computational gain of our proposed algorithm compared to the existing approach, which would though work only for small-scale systems with few number of subcarriers.}
Our numerical results with practical simulation setup show promising results by achieving high spectral efficiency.



\subsection{Organization and Notation:}

\textit{Organization:} The paper is organized as follows. After a brief system model, Section II presents the formulation of rate maximization problem. Section III describes the proposed solution to the problem. Section IV evaluates the performance of our proposed algorithm by assuming practical simulation setup. Finally, Section V concludes the paper.

\textit{Notation:} Bold-face upper-case letters, e.g., $\mathbf{X}$ are used for matrices. Bold-faced lower-case letters, e.g., $\mathbf{x}$,
 are used for vectors and non-bold lower-case letters, e.g., $x$, are used for scalars. $[\mathbf{X}]_{a,b}$ represents $(a,b)^\text{th}$ entry ($a^\text{th}$ row, $b^\text{th}$ column) of the matrix $\mathbf{X}$. Hermitian transpose and normal transpose of the vector $\mathbf{x}$ are denoted by $\mathbf{x}^{H}$, and $\mathbf{x}^{T}$, respectively.  $\mathbb{C}$ is the set of all complex numbers and $\mathbb{R}$ is the set for all real numbers. $ \frac{\partial f(x)}{\partial x} $ is the derivative of function $f(\cdot)$ with respect to its variable $x$. Finally, $\mathbb{E} (X)$ is the expectation operator applied to any random variable $X$.


\section{System Model and Problem Formulation}

\begin{figure*}[t]
    \centering
    \includegraphics[width=0.55 \textwidth]{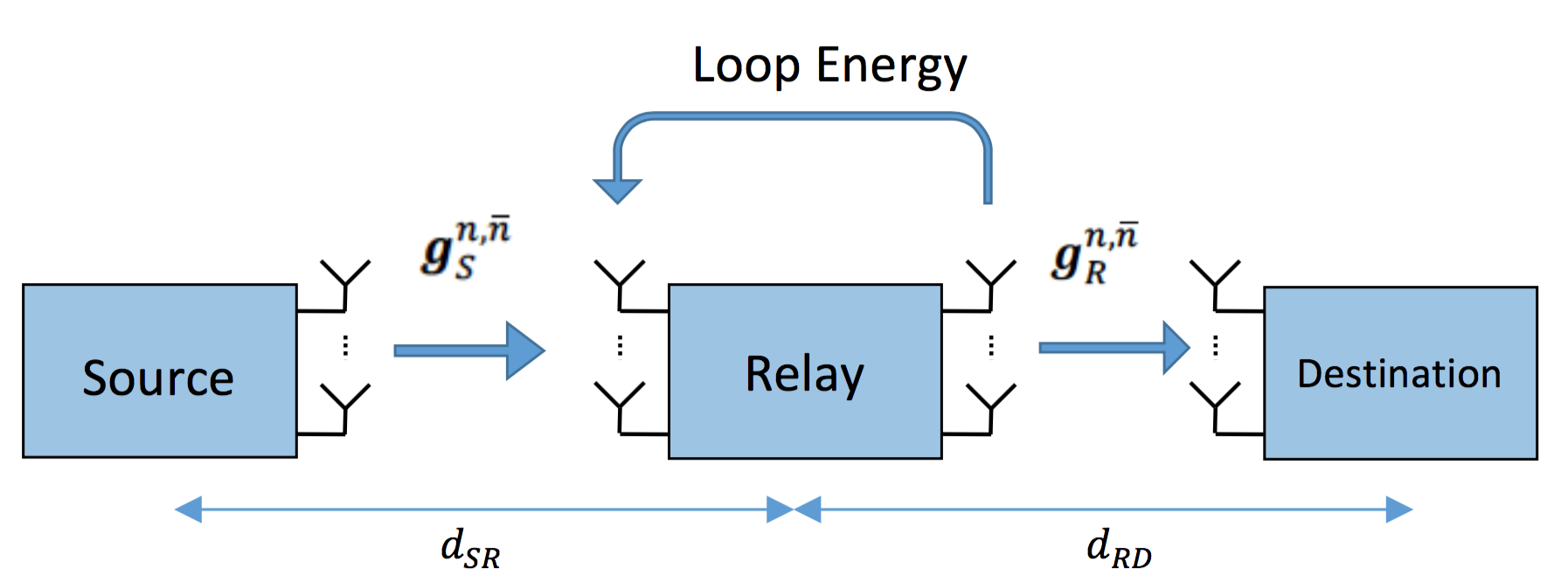}
  \caption{Wireless-powered relaying with self-energy recycling.}
  \label{sys_mod}
\end{figure*}

\begin{figure*}[t]
    \centering
    \includegraphics[width=0.55 \textwidth]{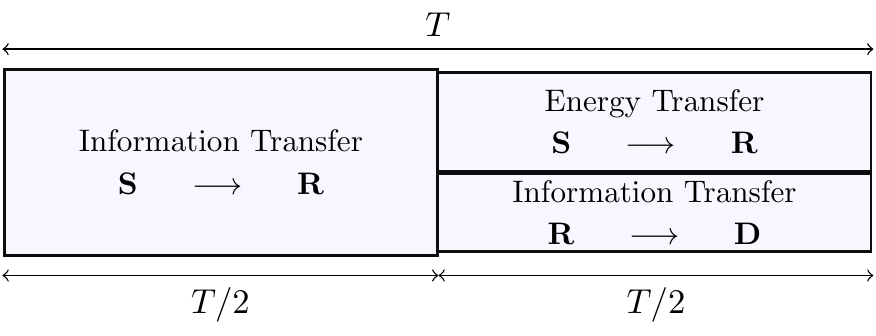}
  \caption{Protocol for full-duplex wireless-powered relaying.}
  \label{BD}
\end{figure*}

Consider a cooperative communication system in which an energy constrained relay node assists the communication between a source node and a destination node. We assume that the destination node is out-of-sight from the source and there is no direct communication link between them. In addition, we assume that the relay node is capable of operating in full-duplex mode and all nodes are equipped with multiple antennas ($N$ antennas), whereas the full-duplex relay node has got $N$ separate antennas at its transmitter and receiver, respectively. As shown in Fig. \ref{sys_mod}, let us denote the distance between the source and the relay node by $d_\text{SR}$ and that between the relay and the destination by $d_\text{RD}$.

As shown in Fig. \ref{BD}, the communication from the source node to the destination node takes place in two-phases. During the first phase, the source sends the information to the relay node. During the second phase, the source sends an energy signal to the relay node and at the same time, the full-duplex relay node forwards the source information to the destination. Note that during the second stage, the relay harvests energy by using not only the dedicated wireless energy signal from the source but also from its own transmission. In other words, the self-interfering link at the relay enables the self-energy recycling.

{Due to multipath propagation, the MIMO channel from the source-to-relay node and from the relay-to-destination node is frequency selective. To combat such multipath channel distortion, multi-carrier modulation, i.e., OFDM is employed for transmitting the information-bearing signal from the source to the relay in the first phase
and for forwarding it by the relay to the destination in the second phase. As such the FD relay can recycle energy 
on sub-carrier base in the second phase. To gain this opportunity of EH on sub-carriers, OFDM is also employed 
for transferring the energy from the source to the relay in the second stage.} 
Let the $L$-tap multipath channel from the $n$th antenna at the source to the $\bar{n}$th antenna at the relay be denoted by $\mathbf{g}_S^{n,\bar{n}} \in \mathbb{C}^{L}$ and that from the $n$th antenna at the relay to the $\bar{n}$th antenna at the destination be denoted by $\mathbf{g}_R^{n,\bar{n}} \in \mathbb{C}^{L}$, $\forall$ $n,  \bar{n} \in \{1,\hdots,N\}$. To ensure meaningful energy harvesting at the relay, the first channel tap of $\mathbf{g}_S^{n,\bar{n}}$ is assumed to be Rician distributed, while the remaining channel taps of $\mathbf{g}_S^{n,\bar{n}}$ and all the channel taps of $\mathbf{g}_R^{n,\bar{n}}$ follow Rayleigh fading. In addition, we have also incorporated the effect of large scale fading within the channel gains $\mathbf{g}_S^{n,\bar{n}}$ and $\mathbf{g}_R^{n,\bar{n}}$. This paper assumes that full channel state information is available by some high-performing channel estimation mechanism and a central processing unit accesses that information to optimize resource allocation.

Let the system bandwidth $B$ be divided into $K$ sub-carriers. Assuming perfect timing and carrier frequency synchronization, the received vector $\tilde{\mathbf{y}}_{R,k} \in \mathbb{C}^{N}$, after the fast Fourier transform (FFT), on sub-carrier $k \in \{1,2,\hdots,K\}$ at the relay is given by
\begin{align}\label{eq:tyR}
     \tilde{\mathbf{y}}_{R,k} = \mathbf{H}_{S,k} \tilde{\mathbf{s}}_{ID,k} +  \tilde{\mathbf{w}}_{R,k},
\end{align}
where
\begin{itemize}
\item $\mathbf{H}_{S,k} \in \mathbb{C}^{N \times N}$ is the MIMO channel matrix between the source and the relay node on subcarrier $k$, such that any element of $\mathbf{H}_{S,k}$, e.g., ($n,\bar{n}$)th element, $[\mathbf{H}_{S,k}]_{n,\bar{n}}$, represents the $k$th sub-channel between the $n$th transmit antenna of the source and the $\bar{n}$th receive antenna of the relay, i.e., we can obtain $[\mathbf{H}_{S,k}]_{n,\bar{n}}$, $\forall$, $k = \{1,\hdots,K\}$ by evaluating the $K$-point FFT of $\mathbf{g}_S^{n,\bar{n}}$,
\item  $\tilde{\mathbf{s}}_{ID,k} = \boldsymbol\Psi_k \mathbf{s}_{ID,k}$, $\boldsymbol\Psi_k \in \mathbb{C}^{N \times N}$ is the transmit precoding matrix on sub-carrier $k$, $\mathbf{s}_{ID,k} \in \mathbb{C}^{N}$ is the modulated information data (ID) on sub-carrier $k$,
\item $\tilde{\mathbf{w}}_{R,k} \in \mathbb{C}^{N}$ is the additive zero-mean Gaussian noise with covariance $\mathbfcal{R}_R = \sigma_R \mathbf{I}_{N}$, $\sigma_R$ is the noise variance at each receive antenna and $\mathbf{I}_{N}$ is the $N \times N$ identity matrix.\footnote{FFT operation at the receiver does not change the noise covaraince.}
\end{itemize}

During the second phase, the relay processes the received signal vector on subcarrier $k$ by a processing matrix $\mathbf{F}_{k} \in \mathbb{C}^{N \times N}$ and forwards the source information to the destination. The received vector $\tilde{\mathbf{y}}_{D,k} \in \mathbb{C}^{N}$, after FFT, on sub-carrier $k \in \{1,2,\hdots,K\}$ at the destination is given by
\begin{align}\label{eq:tyD}
     \tilde{\mathbf{y}}_{D,k} = \mathbf{H}_{R,k} \mathbf{F}_{k} \left( \mathbf{H}_{S,k} \tilde{\mathbf{s}}_{ID,k} +  \tilde{\mathbf{w}}_{R,k} \right) +  \tilde{\mathbf{w}}_{D,k}
\end{align}
where
\begin{itemize}
\item $\mathbf{H}_{R,k} \in \mathbb{C}^{N \times N}$ is the MIMO channel matrix between the relay and the destination node on subcarrier $k$, such that any element of $\mathbf{H}_{R,k}$, e.g., ($n,\bar{n}$)th element, $[\mathbf{H}_{R,k}]_{n,\bar{n}}$, represents the $k$th sub-channel between the $n$th transmit antenna of the relay and the $\bar{n}$th receive antenna of the destination, i.e., we can obtain $[\mathbf{H}_{R,k}]_{n,\bar{n}}$, $\forall$, $k = \{1,\hdots,K\}$ by evaluating the $K$-point FFT of $\mathbf{g}_R^{n,\bar{n}}$,
\item $\tilde{\mathbf{w}}_{D,k} \in \mathbb{C}^{N}$ is the additive zero-mean Gaussian noise with covariance $\mathbfcal{R}_D = \sigma_D \mathbf{I}_{N}$, and $\sigma_D$ is the noise variance at each receive antenna of the destination.
\end{itemize}

\begin{table*}[ht]
\centering
{\color{black}
\caption{List of frequently used symbols with their definitions} 

\begin{tabular}{ |c|c| }
  \hline
  Symbols & Definition \\ \hline
  $h_{S,k,n}$ & channel power grain from source-to-relay over $k$th sub-carrier and $n$th parallel decomposed SISO channel \\
  $h_{R,k,n}$ & channel power grain from relay-to-destination over $k$th sub-carrier and $n$th parallel decomposed SISO channe\\
  $p_{R,k,n}$ & power allocated to the relay to forward the source information over $k$th subcarrier and $n$th transmit antenna  \\
  $p_{ID,k,n}$ & power allocated to the source to transmit source information data over $k$th subcarrier and $n$th transmit antenna  \\
  $p_{EH,k,n}$ & power allocated to the source to transmit energy signal to the relay over $k$th subcarrier and $n$th transmit antenna  \\
  \hline
\end{tabular}
\label{table:symbols} }
\end{table*}

Without loss of generality, we assume that the channel matrices, $\mathbf{H}_{S,k}$ and $\mathbf{H}_{R,k}$, are nonsingular and their singular value decomposition (SVD) is given by
\begin{subequations} \label{chan_decomp}
\begin{align}
\mathbf{H}_{S,k} &= \mathbf{V}_{S,k} \boldsymbol\Lambda_{S,k} \mathbf{U}_{S,k} \\
\mathbf{H}_{R,k} &= \mathbf{V}_{R,k} \boldsymbol\Lambda_{R,k} \mathbf{U}_{R,k}
\end{align}
\end{subequations}
where
\begin{subequations}\label{eq:Lambda}
\begin{align}
 \boldsymbol\Lambda_{S,k} &= \text{diag} \{ \sqrt{h_{S,k,n}} \}_{n=1}^N, \\
  \boldsymbol\Lambda_{R,k} &= \text{diag} \{ \sqrt{h_{R,k,n}} \}_{n=1}^N,
\end{align}
\end{subequations}
\begin{itemize}
\item $\mathbf{V}_{S,k}$, $\mathbf{V}_{R,k}$, $\mathbf{U}_{S,k}$, and $\mathbf{U}_{R,k}$ are unitary matrices of dimension $N \times N$, and
\item $\{h_{S,k,n}\}_{n=1}^N$ and $\{h_{R,k,n}\}_{n=1}^N$ are the eigenvalues of $\mathbf{H}_{S,k} \mathbf{H}_{S,k}^H$ and $ \mathbf{H}_{R,k}^H \mathbf{H}_{R,k}$, respectively. The factors $\sqrt{h_{S,k,n}} \in \mathbb{R}$ and $\sqrt{h_{R,k,n}} \in \mathbb{R}$  can be seen as channel gains from the source-to-relay and from the relay-to-destination over $k$th subcarrier and $n$th parallel decomposed SISO channel, $\forall$, $n \in \{1,\hdots,N\}$.
\end{itemize}

Based on the channel decomposition in \eqref{chan_decomp}, the precoding matrix and the relay processing matrix can be set as follows:
\begin{subequations}\label{eq:Psi_F}
\begin{align}
\boldsymbol\Psi_k &= \mathbf{U}_{S,k}^H \boldsymbol\Gamma_S \\
\mathbf{F}_{k} &= \mathbf{V}_{R,k}^H \boldsymbol\Gamma_R \mathbf{V}_{S,k},
\end{align}
\end{subequations}
where
\begin{subequations}\label{eq:Gamma}
\begin{align}
\boldsymbol\Gamma_S & \triangleq  \text{diag} \{ \sqrt{p_{ID,k,n}} \}_{n=1}^N,  \\
\boldsymbol\Gamma_R & \triangleq  \text{diag} \{ \sqrt{\zeta_{k,n}} \sqrt{p_{R,k,n}} \}_{n=1}^N,
\end{align}
\end{subequations}
\begin{itemize}
\item $p_{ID,k,n}$ is the power allocated to transmit the source information data $s_{ID,k,n}$ over subcarrier $k$ and $n$th transmit antenna, $\forall$, $n \in \{1,\hdots,N\}$,
\item $p_{R,k,n}$ is the power allocated to the relay to forward the source information over subcarrier $k$ and $n$th transmit antenna, and
\item $\zeta_{k,n}$ is the scaling (normalization) factor at the relay to ensure that the signal forwarded, over subcarrier $k$ and $n$th  transmit antenna, carries the power $p_{R,k,n}$,
\end{itemize}

The list of frequently used symbols with their definition is provided in Table \ref{table:symbols}. Using the channel and matrices decomposition in \eqref{chan_decomp} and \eqref{eq:Psi_F}, the received signal vector at the relay, $\tilde{\mathbf{y}}_{R,k}$ in \eqref{eq:tyR}, can be written as
\begin{align}\label{eq:yR}
\mathbf{y}_{R,k} = \boldsymbol\Lambda_{S,k} \boldsymbol\Gamma_S \mathbf{s}_{ID,k} + \mathbf{w}_{R,k}
\end{align}
where
\begin{itemize}
\item $\mathbf{y}_{R,k} = \mathbf{V}_{S,k}^H \tilde{\mathbf{y}}_{R,k}$, and
\item $\mathbf{w}_{R,k} = \mathbf{V}_{S,k}^H \tilde{\mathbf{w}}_{R,k}$ is a noise vector with zero mean and covariance = $\mathbf{V}_{S,k}^H \mathbfcal{R}_R \mathbf{V}_{S,k} $, which is equal to the covariance $\mathbfcal{R}_R = \sigma_R \mathbf{I}_{N}$ due to unitary matrix $ \mathbf{V}_{S,k} $.
\end{itemize}

Similarly, the received signal vector at the destination, $\tilde{\mathbf{y}}_{D,k}$ in \eqref{eq:tyD}, can be written as
\begin{align}\label{eq:yD}
\mathbf{y}_{D,k} = \boldsymbol\Lambda_{R,k} \boldsymbol\Gamma_R \boldsymbol\Lambda_{S,k} \boldsymbol\Gamma_S \mathbf{s}_{ID,k} +  \boldsymbol\Lambda_{R,k} \boldsymbol\Gamma_R \mathbf{w}_{R,k}  + \mathbf{w}_{D,k}
\end{align}
where
\begin{itemize}
\item $\mathbf{y}_{D,k} = \mathbf{U}_{R,k}^H \tilde{\mathbf{y}}_{D,k}$, and
\item $\mathbf{w}_{D,k} = \mathbf{U}_{R,k}^H \tilde{\mathbf{w}}_{D,k}$ is a noise vector with zero mean and covariance = $\mathbf{U}_{R,k}^H \mathbfcal{R}_D \mathbf{U}_{R,k} $, which is equal to the covariance $\mathbfcal{R}_D = \sigma_D \mathbf{I}_{N}$ due to unitary matrix $ \mathbf{U}_{R,k} $.
\end{itemize}

Using \eqref{eq:Lambda} and \eqref{eq:Gamma} and the fact that the MIMO channel on the subcarrier $k$ has been decomposed to $N$ parallel SISO channels, the received signal vector, $\mathbf{y}_{R,k}$ in \eqref{eq:yR}, can be further simplified as
\begin{equation}\label{zz15.1}
y_{R,k,n}=\sqrt{h_{S,k,n}}\sqrt{p_{ID,k,n}}s_{ID,k,n}+\sqrt{\sigma_R}w_{R,k,n},
\end{equation}
where $y_{R,k,n}$ is the received signal at the relay (during the first stage) on the $k$th subcarrier and the $n$th decomposed SISO channel, $s_{ID,k,n}$ is the information data on the $k$th subcarrier and the $n$th transmit antenna such that $\mathbb{E}(|s_{ID,k,n}|^2)=1$, $w_{R,k,n}$ is the normalized noise, such that $\mathbb{E}(|w_{R,k,n}|^2=1)$, $p_{ID,k,n}$ is defined below \eqref{eq:Gamma}, and $\sqrt{h_{S,k,n}}\in \mathbb{R}$ is defined below \eqref{eq:Lambda}.

Similarly, the received signal vector, $\mathbf{y}_{D,k}$ in \eqref{eq:yD}, can be further simplified as
\ifCLASSOPTIONpeerreview
\begin{equation}\label{zz15.2}
y_{D,k,n}=\sqrt{h_{R,k,n}}\sqrt{\frac{p_{R,k,n}}{h_{S,k,n}p_{ID,k,n}+\sigma_R}}
\left(\sqrt{h_{S,k,n}p_{ID,k,n}}s_{ID,k,n}+\sqrt{\sigma_R}w_{R,k,n} \right)+\sqrt{\sigma_D}w_{D,k,n},
\end{equation}
\else
\begin{align}\label{zz15.2}
y_{D,k,n} &=\sqrt{h_{R,k,n}}\sqrt{\frac{p_{R,k,n}}{h_{S,k,n}p_{ID,k,n}+\sigma_R}} \notag \\ & \hspace{0.5cm} \times
\left(\sqrt{h_{S,k,n}p_{ID,k,n}}s_{ID,k,n} +\sqrt{\sigma_R}w_{R,k,n} \right) \notag \\ & \hspace{4cm} +\sqrt{\sigma_D}w_{D,k,n},
\end{align}
\fi
where $y_{D,k,n}$ is the received information data at the destination on the $k$th subcarrier and the $n$th decomposed SISO channel, $w_{D,k,n}$ is the normalized noise, such that $\mathbb{E}(|w_{D,k,n}|^2=1)$, $p_{R,k,n}$ is defined below \eqref{eq:Gamma}, and $\sqrt{h_{R,k,n}}\in \mathbb{R}$ is defined below \eqref{eq:Lambda}. In \eqref{zz15.2},
$\zeta_{k,n} = \frac{p_{R,k,n}}{h_{S,k,n}p_{ID,k,n}+\sigma_R}$ is the relay power normalization factor.

Using \eqref{zz15.2}, the SINR at the destination is given by
\ifCLASSOPTIONpeerreview
\[
\begin{array}{lll}
\text{SINR}_D &=& \ds\frac{h_{R,k,n}p_{R,k,n}h_{S,k,n}p_{ID,k,n}}{(h_{S,k,n}p_{ID,k,n}+\sigma_R)
\left(\frac{h_{R,k,n}p_{R,k,n}\sigma_R}{h_{S,k,n}p_{ID,k,n}+\sigma_R}+\sigma_D \right)}\\[0.5cm]
&=&\ds\frac{h_{R,k,n}p_{R,k,n}h_{S,k,n}p_{ID,k,n}}{
h_{R,k,n}p_{R,k,n}\sigma_R+ (h_{S,k,n}p_{ID,k,n}+\sigma_R)\sigma_D}\\[0.4cm]
&=&\ds\frac{(h_{S,k,n}/\sigma_R)p_{ID,k,n}(h_{R,k,n}/\sigma_D)p_{R,k,n}}{
1+  (h_{S,k,n}/\sigma_R)p_{ID,k,n}+ (h_{R,k,n}/\sigma_D)p_{R,k,n}}.
\end{array}
\]
\else
\[
\begin{array}{ll}
\text{SINR}_D  \hspace{-0.2cm} &= \ds\frac{h_{R,k,n}p_{R,k,n}h_{S,k,n}p_{ID,k,n}}{(h_{S,k,n}p_{ID,k,n}+\sigma_R)
\left(\frac{h_{R,k,n}p_{R,k,n}\sigma_R}{h_{S,k,n}p_{ID,k,n}+\sigma_R}+\sigma_D \right)}\\[0.5cm]
\hspace{-0.2cm}&=\ds\frac{h_{R,k,n}p_{R,k,n}h_{S,k,n}p_{ID,k,n}}{
h_{R,k,n}p_{R,k,n}\sigma_R+ (h_{S,k,n}p_{ID,k,n}+\sigma_R)\sigma_D}\\[0.4cm]
\hspace{-0.2cm}&=\ds\frac{(h_{S,k,n}/\sigma_R)p_{ID,k,n}(h_{R,k,n}/\sigma_D)p_{R,k,n}}{
1+  (h_{S,k,n}/\sigma_R)p_{ID,k,n}+ (h_{R,k,n}/\sigma_D)p_{R,k,n}}.
\end{array}
\]
\fi
During the second stage, the source also transmits dedicated energy signal to the relay with power $p_{EH,k,n}$ and the relay also receives interference from its own transmission. Thus, the received signal at the relay during the second stage, $y_{EH,k,n}$, is given by\footnote{The cyclic prefix in OFDM signaling is discarded at the receiver to eliminate
the inter-symbol interference in information transmission. However, it is subject to EH in energy transfer but its power is very small compared with the whole OFDM signal and thus can be neglected}
\ifCLASSOPTIONpeerreview
\begin{equation}\label{zz15.1EH}
y_{EH,k,n}=\sqrt{h_{S,k,n}}\sqrt{p_{EH,k,n}}s_{EH,k,n}+ \sqrt{h_{LI,k,n}}  \frac{y_{D,k,n} {\color{black} - \sqrt{\sigma_D}w_{D,k,n} } }{\sqrt{h_{R,k,n}}} +  \sqrt{\sigma_R}w_{R,k,n},
\end{equation}
\else
\begin{align}\label{zz15.1EH}
y_{EH,k,n} &=\sqrt{h_{S,k,n}}\sqrt{p_{EH,k,n}}s_{EH,k,n} \notag \\ & \hspace{0.5cm} + \sqrt{h_{LI,k,n}}  \frac{y_{D,k,n} {\color{black} - \sqrt{\sigma_D}w_{D,k,n} } }{\sqrt{h_{R,k,n}}} + \sqrt{\sigma_R}w_{R,k,n},
\end{align}
\fi
where $s_{EH,k,n}$ is the energy signal transmitted over the $k$th subcarrier and from the $n$th antenna of the source, $\sqrt{h_{LI,k,n}}$ is the self-loop interference channel at the full-duplex relay, such that $h_{LI,k,n} \triangleq \gamma_{LI} \tilde{h}_{LI,k,n}$, $\gamma_{LI}$ is the self-loop path gain, and $\sqrt{\tilde{h}_{LI,k,n}} = 1$ is the normalized self-loop interference channel.

Using \eqref{zz15.1EH}, the power harvested by the relay during the second stage over sub-carrier $k$ and $n$th decomposed SISO channel is given by
\begin{align}\label{eq:EH}
e_{k,n}=\eta \left( h_{S,k,n} p_{EH,k,n}+\gamma_{LI} p_{R,k,n} \right) ,
\end{align}
where $\eta$ is the energy harvesting efficiency. Note that in \eqref{eq:EH}, the noise factors containing $\sigma_D$ and $\sigma_R$ have been ignored as they result in negligible harvested energy. In order to ensure that the total transmitted power by the relay does not exceed the total harvested power, the following constraint is imposed:
\begin{equation}\label{zz15.3}
\sum_{k=1}^K\sum_{n=1}^Np_{R,k,n}\leq \sum_{k=1}^K\sum_{n=1}^Ne_{k,n}.
\end{equation}
By setting
\begin{align}
a_{k,n}= \frac{h_{S,k,n}}{\sigma_R}, \ \ \
b_{k,n}=\frac{h_{R,k,n}}{\sigma_D},
\end{align}
the sum-rate maximization problem at the destination can be formulated as
\ifCLASSOPTIONpeerreview
\begin{subequations}\label{p1}
\begin{eqnarray}
\max_{(\mathbf{p}_{ID}, \mathbf{p}_{EH}, \mathbf{p}_{R})}\ f(\mathbf{p}_{ID}, \mathbf{p}_R)\triangleq \sum_{k=1}^K\sum_{n=1}^N\ln
\left(1+\frac{a_{k,n}p_{ID,k,n}b_{k,n}p_{R,k,n}}{1+a_{k,n}p_{ID,k,n}+b_{k,n}p_{R,k,n}} \right)\ \label{p1a}\\
\quad\st\quad \ds\sum_{k=1}^K\sum_{n=1}^N(p_{ID,k,n}+p_{EH,k,n})\leq P,\label{p1b} \\ p_{ID,k,n}\geq 0, p_{EH,k,n}\geq 0, p_{R,k,n}\geq 0,\label{p1b2} \\
\sum_{k=1}^N\sum_{n=1}^Np_{R,k,n}\leq \eta\sum_{k=1}^N\sum_{n=1}^N \left(\sigma_Ra_{k,n}p_{EH,k,n}+\gamma_{LI}p_{R,k,n}\right),
\label{p1c}
\end{eqnarray}
\end{subequations}
where
\else
\begin{subequations}\label{p1}
\begin{align}
& \max_{(\mathbf{p}_{ID}, \mathbf{p}_{EH}, \mathbf{p}_{R})}\  f(\mathbf{p}_{ID}, \mathbf{p}_R) \notag \\  & \hspace{1cm} \triangleq  \sum_{k=1}^K\sum_{n=1}^N\ln
\left(1+\frac{a_{k,n}p_{ID,k,n}b_{k,n}p_{R,k,n}}{1+a_{k,n}p_{ID,k,n}+b_{k,n}p_{R,k,n}} \right)
\ \label{p1a}\\
\quad & \st  \quad \ds\sum_{k=1}^K\sum_{n=1}^N(p_{ID,k,n}+p_{EH,k,n})\leq P,\label{p1b} \\ & \hspace{1cm}  p_{ID,k,n}\geq 0, p_{EH,k,n}\geq 0, p_{R,k,n}\geq 0, \label{p1b2}  \\
&  \sum_{k=1}^N\sum_{n=1}^Np_{R,k,n}\leq \eta\sum_{k=1}^N\sum_{n=1}^N \left(\sigma_Ra_{k,n}p_{EH,k,n}+\gamma_{LI}p_{R,k,n}\right),
\label{p1c}
\end{align}
\end{subequations}
where
\fi
$\mathbf{p}_{ID}\triangleq \{p_{ID,k,n}:\ k=1,\dots, K; n=1,\dots, N\}$, and similarly,
 $\mathbf{p}_{EH}\triangleq \{p_{EH,k,n}:\ k=1,\dots, K; n=1,\dots, N\}$ and $ \mathbf{p}_{R}
 \triangleq \{  p_{R,k,n}:\  k=1,\dots, K; n=1,\dots, N\}$.

Apart from the non-concave objective function \eqref{p1a}, the main issue is the high dimension of (\ref{p1}).
Since the number $K$ of sub-carriers is up to $4096$, the number of optimization variables in (\ref{p1}) can easily
exceed ten thousands, making (\ref{p1}) large-scale nonconvex optimization problem.
{\emph {Our main goal to develop a method that admits a closed-form solution}} at each iteration.

\section{Proposed Solution}
In this section, we propose a solution to solve the non-convex optimization problem \eqref{p1}. In what follows,
define
\begin{equation}\label{def1}
f_{k,n}(p_{ID,k,n}p_{R,k,n}) \hspace{-0.05cm} \triangleq \hspace{-0.05cm} \ln\left(1 \hspace{-0.1cm} + \hspace{-0.1cm} \frac{a_{k,n}p_{ID,k,n}b_{k,n}p_{R,k,n}}{1+a_{k,n}p_{ID,k,n}+b_{k,n}p_{R,k,n}} \right)
\end{equation}
and decompose  it as
\ifCLASSOPTIONpeerreview
\begin{align}
f_{k,n}(p_{ID,k,n}p_{R,k,n})= \ln(1+a_{k,n}p_{ID,k,n})+\ln(1+b_{k,n}p_{R,k,n})
-\ln(1+a_{k,n}p_{ID,k,n}+b_{k,n}p_{R,k,n}).\label{object}
\end{align}
\else
\begin{align}
f_{k,n}(p_{ID,k,n}p_{R,k,n}) &= \ln(1+a_{k,n}p_{ID,k,n}) \notag \\ &+\ln(1+b_{k,n}p_{R,k,n}) \notag \\
&-\ln(1+a_{k,n}p_{ID,k,n}+b_{k,n}p_{R,k,n}).\label{object}
\end{align}
\fi
The first two terms in (\ref{object}) are concave while the last term is convex, so the objective function (\ref{p1a})
is a d.c. (difference of two convex) function \cite{Tuybook}. Thus by linearizing the last term
while keeping the first two terms (\ref{object}) one can easily obtain a lower bounding concave approximation of the
objective function (\ref{p1a}), leading to d.c. iterations (DCI) \cite{KTN12}. Although the constraints (\ref{p1b})-(\ref{p1c}) are polytopic (linear), such DCI is not practical because
the dimension of the convex optimization sub-problem in DCI is too large.  For example, even with only $K=64$ subcarriers and $N=4$ antennas, it takes more than half an hour for DCI on a core-i$5$ machine ($8$ GB RAM) using MATLAB  and CVX solver.

We now develop a new lower bounding concave approximation for the objective function (\ref{p1a}) as well as
a new inner convex approximation for the polytopic constraints (\ref{p1b})-(\ref{p1c}), which lead to a closed-form
of the optimal solution at each iteration.

Let $(\mathbf{p}_{ID}^{(\kappa)}, \mathbf{p}_{EH}^{(\kappa)}, \mathbf{p}_R^{(\kappa)})$ be a feasible point for
(\ref{p1}) that is found from the $(\kappa-1)$th iteration.

\subsection{Lower bounding concave approximation for the objective function (\ref{p1a}) at the $\kappa$th iteration}
By using the inequality (\ref{ap1}) in the appendix:
\ifCLASSOPTIONpeerreview
\begin{equation}\label{obj1}
\ln(1+a_{k,n}p_{ID,k,n})\geq \ln(1+a_{k,n}p^{(\kappa)}_{ID,k,n})+\frac{a_{k,n}p^{(\kappa)}_{ID,k,n}}{1+a_{k,n}p^{(\kappa)}_{ID,k,n}}\left(
\ln p_{ID,k,n}-\ln p^{(\kappa)}_{ID,k,n}
\right)
\end{equation}
\else
\begin{align}\label{obj1}
\ln(1+a_{k,n}p_{ID,k,n}) & \geq \ln(1+a_{k,n}p^{(\kappa)}_{ID,k,n}) \notag \\ & \hspace{-0.8cm} +\frac{a_{k,n}p^{(\kappa)}_{ID,k,n}}{1+a_{k,n}p^{(\kappa)}_{ID,k,n}}\left(
\ln p_{ID,k,n}-\ln p^{(\kappa)}_{ID,k,n}
\right)
\end{align}
\fi
and
\ifCLASSOPTIONpeerreview
\begin{equation}\label{obj2}
\ln(1+b_{k,n}p_{R,k,n})\geq \ln(1+b_{k,n}p^{(\kappa)}_{R,k,n})+\frac{b_{k,n}p^{(\kappa)}_{R,k,n}}{1+b_{k,n}p^{(\kappa)}_{R,k,n}}\left(
\ln p_{R,k,n}-\ln p^{(\kappa)}_{R,k,n}
\right),
\end{equation}
\else
\begin{align}\label{obj2}
\ln(1+b_{k,n}p_{R,k,n}) & \geq \ln(1+b_{k,n}p^{(\kappa)}_{R,k,n}) \notag \\ & \hspace{-0.8cm}+\frac{b_{k,n}p^{(\kappa)}_{R,k,n}}{1+b_{k,n}p^{(\kappa)}_{R,k,n}}\left(
\ln p_{R,k,n}-\ln p^{(\kappa)}_{R,k,n}
\right),
\end{align}
\fi
while by using the inequality (\ref{ap2}) in the appendix
\ifCLASSOPTIONpeerreview
\begin{eqnarray}\label{obj3}
\ln(1+a_{k,n}p_{ID,k,n}+b_{k,n}p_{R,k,n})&\leq& \ln(1+a_{k,n}p^{(\kappa)}_{ID,k,n}+b_{k,n}p^{(\kappa)}_{R,k,n})\nonumber\\
&&\ds+\frac{a_{k,n}(p_{ID,k,n}-p^{(\kappa)}_{ID,k,n})+b_{k,n}(p_{R,k,n}-p^{(\kappa)}_{R,k,n}) }{1+a_{k,n}p^{(\kappa)}_{ID,k,n}+b_{k,n}p^{(\kappa)}_{R,k,n}}.
\end{eqnarray}
\else
\begin{align}\label{obj3}
& \ln(1+a_{k,n}p_{ID,k,n}+b_{k,n}p_{R,k,n}) \notag \\
& \hspace{0.3cm} \leq
 \ln(1+a_{k,n}p^{(\kappa)}_{ID,k,n}+b_{k,n}p^{(\kappa)}_{R,k,n})\nonumber\\
& \hspace{0.6cm} \ds+\frac{a_{k,n}(p_{ID,k,n}-p^{(\kappa)}_{ID,k,n})+b_{k,n}(p_{R,k,n}-p^{(\kappa)}_{R,k,n}) }{1+a_{k,n}p^{(\kappa)}_{ID,k,n}+b_{k,n}p^{(\kappa)}_{R,k,n}}.
\end{align}
\fi
Therefore,
\ifCLASSOPTIONpeerreview
\begin{eqnarray}
f_{k,n}(p_{ID,k,n},p_{R,k,n})&\geq&
\fk_{k,n}(p_{ID,k,n},p_{R,k,n})\nonumber\\
&\triangleq& \ds\alphak_{k,n}+\betak_{k,n}\ln p_{ID,k,n}-\chik_{k,n}p_{ID,k,n}+\deltak_{k,n}\ln p_{R,k,n}-\gammak_{k,n}p_{R,k,n},\label{obj4}
\end{eqnarray}
\else
\begin{align}
f_{k,n}(p_{ID,k,n},p_{R,k,n})&\geq
\fk_{k,n}(p_{ID,k,n},p_{R,k,n})\nonumber\\
& \hspace{-0.25cm} \triangleq \ds\alphak_{k,n}+\betak_{k,n}\ln p_{ID,k,n}-\chik_{k,n}p_{ID,k,n} \notag \\ & \hspace{0.5cm} +\deltak_{k,n}\ln p_{R,k,n}-\gammak_{k,n}p_{R,k,n},\label{obj4}
\end{align}
\fi
where
\ifCLASSOPTIONpeerreview
\begin{subequations}
\begin{align}
\alphak_{k,n}&= \ds \ln \left(1+\frac{a_{k,n}\pk_{ID,k,n}b_{k,n}\pk_{R,k,n}}{1+a_{k,n}\pk_{ID,k,n}+b_{k,n}\pk_{R,k,n}}\right)
-\frac{a_{k,n}\pk_{ID,k,n}}{1+a_{k,n}\pk_{ID,k,n}}\ln\pk_{ID,k,n} \notag \\
&\ds-\frac{b_{k,n}\pk_{R,k,n}}{1+b_{k,n}\pk_{R,k,n}}\ln\pk_{R,k,n}+\frac{1}{1+a_{k,n}\pk_{ID,k,n}+b_{k,n}\pk_{R,k,n}}\left(a_{k,n}\pk_{ID,k,n} + b_{k,n}\pk_{R,k,n})\right), \\
\betak &=\ds\frac{a_{k,n}\pk_{ID,k,n}}{1+a_{k,n}\pk_{ID,k,n}} > 0, \\
\chik_{k,n} &=\ds\frac{1}{1+a_{k,n}\pk_{ID,k,n}+b_{k,n}\pk_{R,k,n}}a_{k,n} > 0, \\
\deltak_{k,n} &=\ds\frac{b_{k,n}\pk_{R,k,n}}{1+b_{k,n}\pk_{R,k,n}} > 0, \\
\gammak_{k,n} &=\ds\frac{1}{1+a_{k,n}\pk_{ID,k,n}+b_{k,n}\pk_{R,k,n}}b_{k,n} > 0.
\end{align}
\end{subequations}
\else
\begin{subequations}
\begin{align}
\alphak_{k,n}&= \ds \ln \left(1+\frac{a_{k,n}\pk_{ID,k,n}b_{k,n}\pk_{R,k,n}}{1+a_{k,n}\pk_{ID,k,n}+b_{k,n}\pk_{R,k,n}}\right)
 \notag \\
& \hspace{-0.25cm} \ds -\frac{a_{k,n}\pk_{ID,k,n}}{1+a_{k,n}\pk_{ID,k,n}}\ln\pk_{ID,k,n} -\frac{b_{k,n}\pk_{R,k,n}}{1+b_{k,n}\pk_{R,k,n}}\ln\pk_{R,k,n}  \notag \\
&\hspace{-0.25cm} +\frac{1}{1+a_{k,n}\pk_{ID,k,n}+b_{k,n}\pk_{R,k,n}} \notag \\ & \hspace{2.2cm}  \times \left(a_{k,n}\pk_{ID,k,n} + b_{k,n}\pk_{R,k,n})\right), \\
\betak &=\ds\frac{a_{k,n}\pk_{ID,k,n}}{1+a_{k,n}\pk_{ID,k,n}} > 0, \\
\chik_{k,n} &=\ds\frac{1}{1+a_{k,n}\pk_{ID,k,n}+b_{k,n}\pk_{R,k,n}}a_{k,n} > 0, \\
\deltak_{k,n} &=\ds\frac{b_{k,n}\pk_{R,k,n}}{1+b_{k,n}\pk_{R,k,n}} > 0, \\
\gammak_{k,n} &=\ds\frac{1}{1+a_{k,n}\pk_{ID,k,n}+b_{k,n}\pk_{R,k,n}}b_{k,n} > 0.
\end{align}
\end{subequations}
\fi
Note that $\fk_{k,n}(p_{ID,k,n},p_{R,k,n})$ is a concave function, which matches with $f_{k,n}(p_{ID,k,n},p_{R,k,n})$ at $(p^{(\kappa)}_{ID,k,n},p^{(\kappa)}_{R,k,n})$, i.e. $f_{k,n}(p^{(\kappa)}_{ID,k,n},p^{(\kappa)}_{R,k,n})=
\fk_{k,n}(p^{(\kappa)}_{ID,k,n},p^{(\kappa)}_{R,k,n})$. Consequently,  the function
\[
f^{(\kappa)}(\mathbf{p}_{ID}, \mathbf{p}_R)\triangleq \sum_{k=1}^K\sum_{n=1}^N\fk_{k,n}(p_{ID,k,n},p_{R,k,n})
\]
is a lower bounding concave approximation of $f(\mathbf{p}_{ID}, \mathbf{p}_R)$:
\begin{equation}\label{lw1}
f(\mathbf{p}_{ID}, \mathbf{p}_R)\geq f^{(\kappa)}(\mathbf{p}_{ID}, \mathbf{p}_R)\quad\forall\ (\mathbf{p}_{ID}, \mathbf{p}_R),
\end{equation}
and matches with $f(\mathbf{p}_{ID}, \mathbf{p}_R)$ at $(\mathbf{p}^{(\kappa)}_{ID}, \mathbf{p}^{(\kappa)}_R)$:
\begin{equation}\label{lw2}
f(\mathbf{p}^{(\kappa)}_{ID}, \mathbf{p}^{(\kappa)}_R)= f^{(\kappa)}(\mathbf{p}^{(\kappa)}_{ID}, \mathbf{p}^{(\kappa)}_R).
\end{equation}
\subsection{Inner approximation of polytopic constraints at the $\kappa$th iteration}
Note that $p_{EH,k,n}$ appears only in the polytopic constraints (\ref{p1b}) and (\ref{p1c}). Now, note that
\ifCLASSOPTIONpeerreview
\[
p^2_{EH,k,n}+(p^{(\kappa)}_{EH,k,n})^2-2p_{EH,k,n}p^{(\kappa)}_{EH,k,n}=(p_{EH,k,n}-p^{(\kappa)}_{EH,k,n})^2\geq 0,
\]
\else
\begin{align}
& p^2_{EH,k,n}+(p^{(\kappa)}_{EH,k,n})^2-2p_{EH,k,n}p^{(\kappa)}_{EH,k,n} \notag \\ & \hspace{4cm} =(p_{EH,k,n}-p^{(\kappa)}_{EH,k,n})^2 \notag \\ & \hspace{4cm} \geq 0 \notag
\end{align}
\fi
leading to
\[
p_{EH,k,n}\leq 0.5\left(p_{EH,k,n}/p^{(\kappa)}_{EH,k,n}+p^{(\kappa)}_{EH,k,n}\right),
\]
we innerly approximate the polytopic constraint (\ref{p1b}) by the convex quadratic constraint
\begin{equation}\label{p1be}
\ds\sum_{k=1}^K\sum_{n=1}^N\left(p_{ID,k,n}+0.5\left( \frac{p^2_{EH,k,n}}{\pk_{EH,k,n}}+\pk_{EH,k,n} \right)\right)\leq P.
\end{equation}
Indeed, it is obvious that any feasible point for (\ref{p1be}) is also feasible for (\ref{p1b}).
\subsection{Closed-form solution at the $\kappa$th iteration}
At the $\kappa$th iteration we solve the following  convex optimization problem to generate the next iterative
feasible point $(\mathbf{p}_{ID}^{(\kappa+1)}, \mathbf{p}_{EH}^{(\kappa+1)}, \mathbf{p}_R^{(\kappa+1)})$ for
(\ref{p1}):
\begin{equation}\label{p2}
\max_{(\mathbf{p}_{ID}, \mathbf{p}_{EH}, \mathbf{p}_{R})}\ f^{(\kappa)}(\mathbf{p}_{ID}, \mathbf{p}_R)  \quad \st \quad (\ref{p1c}), (\ref{p1b2}), (\ref{p1be}).
\end{equation}
The Lagrangian of problem \eqref{p2} is given by
\ifCLASSOPTIONpeerreview
\[
\begin{array}{lll}
\clL(\mathbf{p}_{ID},\mathbf{p}_{EH},\mathbf{p}_{R},\lambda_1,\lambda_2)&=&\ds\sum_{k=1}^K\sum_{n=1}^N\fk_{k,n}(p_{ID,k,n},p_{R,k,n})\\
&&\ds-\lambda_1\left(\sum_{k=1}^K\sum_{n=1}^N\left(p_{ID,k,n}+0.5 \left(p^2_{EH,k,n}/\pk_{EH,k,n}+\pk_{EH,k,n}\right)\right)-P  \right)\\
&&\ds-\lambda_2\left(\sum_{k=1}^N\sum_{n=1}^Np_{R,k,n}-\eta\sum_{k=1}^N\sum_{n=1}^N \left(\sigma_Ra_{k,n}p_{EH,k,n}+\gamma_{LI}p_{R,k,n}\right)
\right),
\end{array}
\]
\else
\begin{align}
& \clL(\mathbf{p}_{ID},\mathbf{p}_{EH},\mathbf{p}_{R},\lambda_1,\lambda_2) \notag \\ &= \ds\sum_{k=1}^K\sum_{n=1}^N\fk_{k,n}(p_{ID,k,n},p_{R,k,n}) \notag \\
&\ds-\lambda_1\left(\sum_{k=1}^K\sum_{n=1}^N\left(p_{ID,k,n}+\frac{1}{2} \left( \frac{p^2_{EH,k,n}}{\pk_{EH,k,n}}+\pk_{EH,k,n}\right)\right)-P \hspace{-0.05cm} \right) \notag \\
&\ds-\lambda_2\bigg(\sum_{k=1}^N\sum_{n=1}^Np_{R,k,n}-\eta\sum_{k=1}^N\sum_{n=1}^N \Big(\sigma_Ra_{k,n}p_{EH,k,n} \notag \\ & \hspace{6.5cm} +\gamma_{LI}p_{R,k,n}\Big)
\bigg), \notag
\end{align}
\fi
where $\lambda_1>0$ and $\lambda_2>0$ are the Lagrange multipliers.

The optimal solution of (\ref{p2}) satisfies
\ifCLASSOPTIONpeerreview
\begin{eqnarray}
\ds\frac{\partial \clL(\mathbf{p}_{ID},\mathbf{p}_{EH},\mathbf{p}_{R},\lambda_1,\lambda_2)}{\partial p_{EH,k,n}}=0&\Leftrightarrow&
-\frac{\lambda_1p_{EH,k,n}}{p_{EH,k,n}^{(\kappa)}} +\lambda_2\eta\sigma_Ra_{k,n}=0\label{p3a}\\
&\Leftrightarrow&p_{EH,k,n}=\lambda_2\eta\sigma_Ra_{k,n} p_{EH,k,n}^{(\kappa)} /\lambda_1,\label{p3}
\end{eqnarray}
\else
\begin{align}
& \ds\frac{\partial \clL(\mathbf{p}_{ID},\mathbf{p}_{EH},\mathbf{p}_{R},\lambda_1,\lambda_2)}{\partial p_{EH,k,n}}=0 \notag \\ &\Leftrightarrow
-\frac{\lambda_1p_{EH,k,n}}{p_{EH,k,n}^{(\kappa)}} +\lambda_2\eta\sigma_Ra_{k,n}=0\label{p3a}\\
&\Leftrightarrow p_{EH,k,n}=\lambda_2\eta\sigma_Ra_{k,n} p_{EH,k,n}^{(\kappa)} /\lambda_1,\label{p3}
\end{align}
\fi
and
\ifCLASSOPTIONpeerreview
\begin{eqnarray}\label{p32}
\ds\frac{\partial \clL(\mathbf{p}_{ID},\mathbf{p}_{EH},\mathbf{p}_{R},\lambda_1,\lambda_2)}{\partial p_{ID,k,n}}=0&\Leftrightarrow&
\ds\frac{\betak_{k,n}}{p_{ID,k,n}}-\chik_{k,n}-\lambda_1=0\nonumber\\
&\Leftrightarrow&p_{ID,k,n}=\ds\frac{\betak_{k,n}}{\chik_{k,n}+\lambda_1},
\end{eqnarray}
\else
\begin{align}\label{p32}
& \ds\frac{\partial \clL(\mathbf{p}_{ID},\mathbf{p}_{EH},\mathbf{p}_{R},\lambda_1,\lambda_2)}{\partial p_{ID,k,n}}=0 \notag \\ &\Leftrightarrow
\ds\frac{\betak_{k,n}}{p_{ID,k,n}}-\chik_{k,n}-\lambda_1=0\nonumber\\
&\Leftrightarrow p_{ID,k,n}=\ds\frac{\betak_{k,n}}{\chik_{k,n}+\lambda_1},
\end{align}
\fi
and
\ifCLASSOPTIONpeerreview
\begin{eqnarray}\label{p4}
\ds\frac{\partial \clL(\mathbf{p}_{ID},\mathbf{p}_{EH},\mathbf{p}_{R},\lambda_1,\lambda_2)}{\partial p_{R,k,n}}=0&\Leftrightarrow&
\ds\frac{\deltak_{k,n}}{p_{R,k,n}}-\gammak_{k,n}-\lambda_2(1-\eta\gamma_{LI})=0\nonumber\\
&\Leftrightarrow&p_{R,k,n}=\ds\frac{\deltak_{k,n}}{\gammak_{k,n}+\lambda_2(1-\eta\gamma_{LI})},
\end{eqnarray}
\else
\begin{align}\label{p4}
& \ds\frac{\partial \clL(\mathbf{p}_{ID},\mathbf{p}_{EH},\mathbf{p}_{R},\lambda_1,\lambda_2)}{\partial p_{R,k,n}}=0 \notag \\ &\Leftrightarrow
\ds\frac{\deltak_{k,n}}{p_{R,k,n}}-\gammak_{k,n}-\lambda_2(1-\eta\gamma_{LI})=0\nonumber\\
&\Leftrightarrow p_{R,k,n}=\ds\frac{\deltak_{k,n}}{\gammak_{k,n}+\lambda_2(1-\eta\gamma_{LI})},
\end{align}
\fi
where $\lambda_1>0$ and $\lambda_2>0$ are chosen such that the constraints \eqref{p1b} and \eqref{p1c} are met with equality. Thus,
\ifCLASSOPTIONpeerreview
\begin{eqnarray}
&& \ds\sum_{k=1}^N\sum_{n=1}^N(p_{ID,k,n}+p_{EH,k,n})=P \notag \\
&\Leftrightarrow&
\ds \sum_{k=1}^N\sum_{n=1}^N\left(\frac{\betak_{k,n}}{\chik_{k,n}+\lambda_1}+ \frac{\lambda_2\eta\sigma_Ra_{k,n} p_{EH,k,n}^{(\kappa)} }{\lambda_1} \right)=P,\label{p5}
\end{eqnarray}
\else
\begin{align}
& \ds\sum_{k=1}^N\sum_{n=1}^N(p_{ID,k,n}+p_{EH,k,n})=P \notag \\
&\Leftrightarrow
\ds \sum_{k=1}^N\sum_{n=1}^N\left(\frac{\betak_{k,n}}{\chik_{k,n}+\lambda_1}+ \frac{\lambda_2\eta\sigma_Ra_{k,n} p_{EH,k,n}^{(\kappa)} }{\lambda_1} \right)=P,\label{p5}
\end{align}
\fi
and
\ifCLASSOPTIONpeerreview
\begin{eqnarray}
&& \ds\sum_{k=1}^N\sum_{n=1}^N p_{R,k,n} = \eta \sum_{k=1}^N\sum_{n=1}^N  \left(\sigma_Ra_{k,n}p_{EH,k,n}+\gamma_{LI}p_{R,k,n}\right)  \notag \\ &\Leftrightarrow&
(1 - \eta \gamma_{LI}) \ds \sum_{k=1}^N\sum_{n=1}^N \ds\frac{\deltak_{k,n}}{\gammak_{k,n}+\lambda_2(1-\eta\gamma_{LI})}= \eta \ds\sum_{k=1}^N\sum_{n=1}^N \frac{\lambda_2 \sigma_R  a_{k,m} \eta\sigma_Ra_{k,n} p_{EH,k,n}^{(\kappa)} }{\lambda_1}.\label{p6}
\end{eqnarray}
\else
\begin{align}
& \ds\sum_{k=1}^N\sum_{n=1}^N p_{R,k,n} = \eta \sum_{k=1}^N\sum_{n=1}^N  \left(\sigma_Ra_{k,n}p_{EH,k,n}+\gamma_{LI}p_{R,k,n}\right)  \notag \\ &\Leftrightarrow
(1 - \eta \gamma_{LI}) \ds \sum_{k=1}^N\sum_{n=1}^N \ds\frac{\deltak_{k,n}}{\gammak_{k,n}+\lambda_2(1-\eta\gamma_{LI})} \notag \\ & \hspace{2cm} = \eta \ds\sum_{k=1}^N\sum_{n=1}^N \frac{\lambda_2 \sigma_R  a_{k,m} \eta\sigma_Ra_{k,n} p_{EH,k,n}^{(\kappa)} }{\lambda_1}.\label{p6}
\end{align}
\fi
Simultaneously solving \eqref{p5} and \eqref{p6} will yield the optimal values of $\lambda_2$ and $\lambda_1$. From \eqref{p6}, we can get $\lambda_1$ expressed as a function of $\lambda_2$:
\begin{align}\label{lam1}
\lambda_1 = \frac{\eta \lambda_2}{1 - \eta \sigma_{LI}} \frac{\sum_{k=1}^N\sum_{n=1}^N  \eta \sigma_R^2 a_{k,n}^2 \pk_{EH,k,n} }{\ds \sum_{k=1}^N\sum_{n=1}^N \left(\frac{\deltak_{k,n}}{\gammak_{k,n}+\lambda_2(1-\eta\gamma_{LI})} \right) }.
\end{align}
Substituting this value of $\lambda_1$ into \eqref{p5}, we can solve for $\lambda_2$ and finally get $\lambda_1$ from \eqref{lam1}.

{\bf Remark 1.} In (\ref{obj1}) and (\ref{obj2}) we approximate concave functions by other concave functions, which lead to the simple closed forms (\ref{p32}) and (\ref{p4}) and consequently, the analytical formula (\ref{lam1}) for determining the Lagrange multiplier $\lambda_1$.

{\bf Remark 2.} It can be easily seen that using the polytopic constraint (\ref{p1b}) cannot reveal a closed-form of
$p_{EH,k,n}$ as the derivative of the corresponding Lagrangian with respect to $p_{EH,k,n}$ as in (\ref{p3a})
is independent of  $p_{EH,k,n}$. That is why we need the inner approximation (\ref{p1be}) for (\ref{p1b}) that leads
to the closed-form (\ref{p3}) for $p_{EH,k,n}$.

{\color{black}
{\bf Remark 3.} At the $\kappa$th iteration, the DCI solution can be obtained by iteratively solving the following optimization problem
\begin{align}\label{dc_prob}
\max_{(\mathbf{p}_{ID}, \mathbf{p}_{EH}, \mathbf{p}_{R})}\ & \ln(1+a_{k,n}p_{ID,k,n})+\ln(1+b_{k,n}p_{R,k,n}) \notag \\ &
- \ln(1+a_{k,n}p^{(\kappa)}_{ID,k,n}+b_{k,n}p^{(\kappa)}_{R,k,n})\nonumber\\
& \hspace{-1cm} - \frac{a_{k,n}(p_{ID,k,n}-p^{(\kappa)}_{ID,k,n})+b_{k,n}(p_{R,k,n}-p^{(\kappa)}_{R,k,n}) }{1+a_{k,n}p^{(\kappa)}_{ID,k,n}+b_{k,n}p^{(\kappa)}_{R,k,n}}. \notag \\  &\quad \st \quad (\ref{p1b}), (\ref{p1b2}), (\ref{p1c}),
\end{align}
where the difference between the previously formulated problem \eqref{p2} and the above DCI problem \eqref{dc_prob} is this that in the former, we further approximate the concave functions $\ln(1+a_{k,n}p_{ID,k,n})$ and $\ln(1+b_{k,n}p_{R,k,n})$ by other concave functions to ease the process of finding a closed form solution.}

\subsection{Algorithm and convergence}
The following feasible point $(\mathbf{p}_{EH}^{(0)}, \mathbf{p}_{ID}^{(0)}, \mathbf{p}_{R}^{(0)})$
is taken  as the initial point:
\begin{subequations}\label{eq:ini}
\begin{align}
p_{EH,k,n}^{(0)} &= \frac{0.5 P}{K N} \\
 p_{ID,k,n}^{(0)} &= \frac{0.5 P}{K N} \\
 p_{R,k,n}^{(0)} &=  \frac{\eta h_{S,k,n} p_{EH,k,n}^{(0)}}{ 1 - \eta \gamma_{LI} }.
\end{align}
\end{subequations}
It can be easily seen that all the constraints \eqref{p1b}, \eqref{p1b2}, and \eqref{p1c} are met.
Algorithm \ref{alg1} outlines the steps to solve the rate maximization problem \eqref{p1}.

\begin{algorithm}[t]
\begin{algorithmic}[1]
\protect\caption{Resource Allocation Algorithm for Rate Maximization Problem \eqref{p1}.}
\label{alg1}
\global\long\def\algorithmicrequire{\textbf{Initialization:}}
\REQUIRE  Set $\kappa:=0$. Take the initial feasible point $(\mathbf{p}_{EH}^{(0)}, \mathbf{p}_{ID}^{(0)}, \mathbf{p}_{R}^{(0)})$  by using (\ref{eq:ini}).
\REPEAT
\STATE Solve for $\lambda_1$ and $\lambda_2$ using \eqref{p5} and \eqref{lam1} (the details are given below \eqref{p6}).
\STATE Update $(\mathbf{p}_{EH}^{(\kappa+1)}, \mathbf{p}_{ID,k,n}^{(\kappa+1)}, \mathbf{p}_{R,k,n}^{(\kappa+1)})$ using
  the closed-forms \eqref{p3}, \eqref{p32}, and \eqref{p4}.
\STATE Set $\kappa:=\kappa+1.$
\UNTIL Convergence\\
\end{algorithmic} \end{algorithm}

By (\ref{lw2}),
\[
f(\mathbf{p}^{(\kappa)}_{ID}, \mathbf{p}^{(\kappa)}_R)=f^{(\kappa)}(\mathbf{p}^{(\kappa)}_{ID}, \mathbf{p}^{(\kappa)}_R).
\]
Furthermore,
\[
f^{(\kappa)}(\mathbf{p}^{(\kappa+1)}_{ID}, \mathbf{p}^{(\kappa+1)}_R)>
f^{(\kappa)}(\mathbf{p}^{(\kappa)}_{ID}, \mathbf{p}^{(\kappa)}_R)
\]
because $(\mathbf{p}_{EH}^{(\kappa+1)}, \mathbf{p}_{ID}^{(\kappa+1)}, \mathbf{p}_{R}^{(\kappa+1)})$ is the
optimal solution of (\ref{p2}) while $(\mathbf{p}_{EH}^{(\kappa)}, \mathbf{p}_{ID}^{(\kappa)}, \mathbf{p}_{R}^{(\kappa)})$ is a feasible point for (\ref{p2}).
Therefore, by (\ref{lw1}),
\[
\begin{array}{lll}
f(\mathbf{p}^{(\kappa+1)}_{ID}, \mathbf{p}^{(\kappa+1)}_R)&\geq& f^{(\kappa)}(\mathbf{p}^{(\kappa+1)}_{ID}, \mathbf{p}^{(\kappa+1)}_R)\\
&>&f^{(\kappa)}(\mathbf{p}^{(\kappa)}_{ID}, \mathbf{p}^{(\kappa)}_R)\\
&=&f(\mathbf{p}^{(\kappa)}_{ID}, \mathbf{p}^{(\kappa)}_R),
\end{array}
\]
i.e. $(\mathbf{p}_{EH}^{(\kappa+1)}, \mathbf{p}_{ID}^{(\kappa+1)}, \mathbf{p}_{R}^{(\kappa+1)})$ is a better feasible
point $(\mathbf{p}_{EH}^{(\kappa)}, \mathbf{p}_{ID}^{(\kappa)}, \mathbf{p}_{R}^{(\kappa)})$ for the original nonconvex optimization problem (\ref{p1}). As such, Algorithm \ref{alg1} converges at least to a locally optimal solution of
(\ref{p1}) \cite{Nasir-16-TCOM-A}.

\section{Numerical Results}

\ifCLASSOPTIONpeerreview
\else
\begin{figure*}[t]
    \centering
    \begin{minipage}[h]{0.48\textwidth}
    \centering
    \includegraphics[width=1.01 \textwidth]{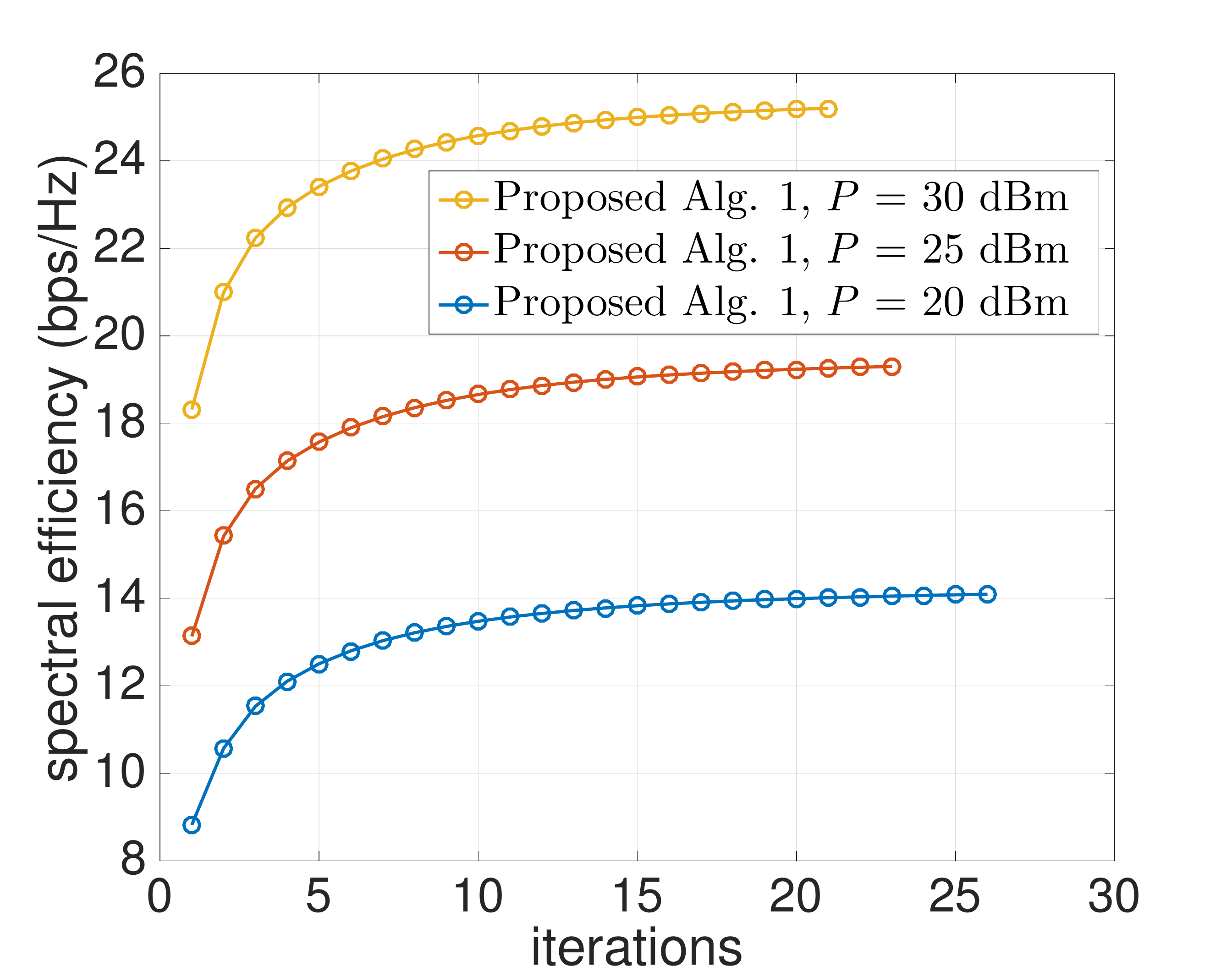}
  \caption{The convergence of Proposed Alg. \ref{alg1}.}
  \label{fig:iter_P}
  \end{minipage}
    \hspace{0.3cm}
    \begin{minipage}[h]{0.48\textwidth}
    \centering
    \includegraphics[width=1.01 \textwidth]{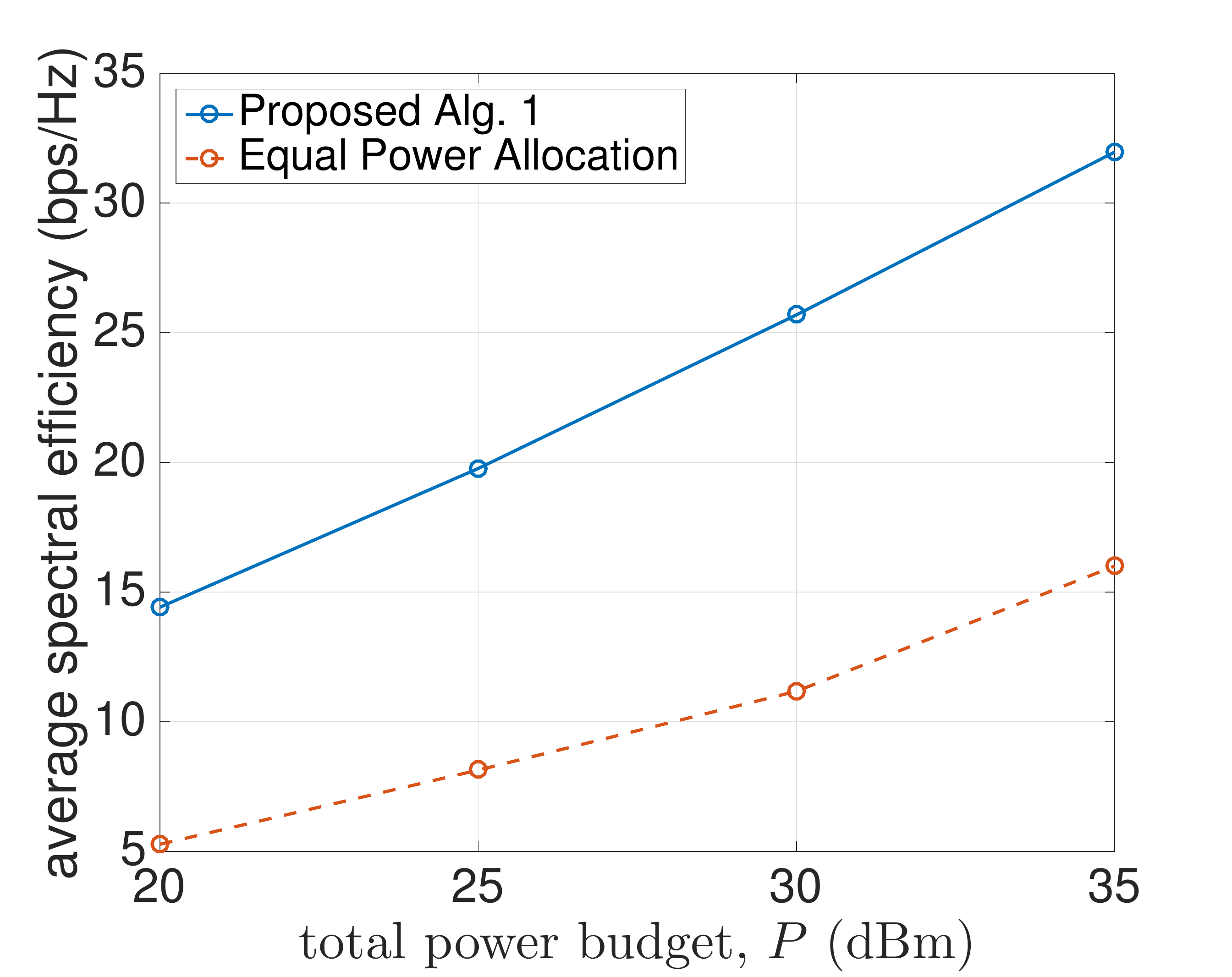}
  \caption{Average spectral efficiency versus the total power budget $P$.}
  \label{fig:r_P}
  \end{minipage}
\end{figure*}
\fi

In this section, we will evaluate the performance of the proposed Alg. \ref{alg1}. We first define the simulation setup. We assume $N = 4$ antennas at each node. To model large scale fading, each spatial $L = 16$-tap source-to-relay channel, $\mathbf{g}_S^{n,\bar{n}}$, between any transmit antenna $n$ and receive antenna $\bar{n}$, follows the path loss model $30 + 10 \beta \log_{10} (d_{SR})$, where the path loss exponent $\beta = 2$. On the other hand, the $L = 16$-tap relay-to-destination channel, $\mathbf{g}_R^{n,\bar{n}}$ is modeled by the path loss model $30 + 10 \beta \log_{10} (d_{RD})$ with the path loss exponent $\beta = 3$ (the path-loss model is consistent with dense nodes deployment and wireless energy harvesting requirements at the relay \cite{Do-17-A-EA}). The distance $d_{SR}$ is set to $10$ meters, while unless specified otherwise, the distance $d_{RD}$ is set to $50$ meters. {\color{black}Note that different values for path-loss exponents have been proposed for different type of channel conditions in the literature too \cite{Liu-17-Dec-A,Ghazanfari-Apr-16-A}, where to ensure meaningful wireless power transfer to energy harvesting node, smaller distance with line-of-sight component and hence, smaller value of path-loss exponent are adopted in the literature \cite{Ghazanfari-Apr-16-A}.}

To model small scale fading, the first channel tap of $\mathbf{g}_S^{n,\bar{n}}$ is assumed to be Rician distributed with Rician factor $\mathcal{K} = 6$ dB, while the remaining channel taps of $\mathbf{g}_S^{n,\bar{n}}$ and all the channel taps of $\mathbf{g}_R^{n,\bar{n}}$ are Rayleigh distributed. To simulate the effect of frequency selectivity in each spatial multipath channel, we assume an exponential power delay profile with root-mean-square delay spread of $\sigma_\text{RMS} = 3 T_s$, for the symbol time $T_s = 1/B$. In addition, the spatial correlation among the MIMO channels are modeled according to Case B of the 3 GPP I-METRA MIMO channel model \cite{Cho-B-10}.

The time-domain channels  $\mathbf{g}_S^{n,\bar{n}}$ and $\mathbf{g}_R^{n,\bar{n}}$ are converted to frequency domain via a $K$-point FFT. We consider $K = 1024$ OFDM subcarriers, unless specified otherwise. The system bandwidth is set to $B = 1$-MHz and the subcarrier bandwidth is $B/K$. This subcarrier bandwidth is quite smaller than the coherence bandwidth of $0.02/\sigma_{RMS}$ to ensure flat fading over each subcarrier. In each sub-channel, the power spectral density of additive white Gaussian noise, $ \frac{\sigma_R }{ B/K }$ and $ \frac{\sigma_D }{ B/K }$ at each antenna is set to $ - 174$ dBm/Hz. The correlation between noise samples from different antennas is set to $0.2$. The carrier frequency is assumed to be $1$ GHz. Unless specified otherwise, the self-loop path loss is set to $ \frac{1}{\gamma_{LI}} = 10$ dB, {\color{black} which is justified since there is no source-to-relay information transfer during the second communication phase and there is no need of explicit self-interference attenuation at the relay because energy in the self-interference channel is recycled by the relay via wireless energy harvesting \cite{Chen-18-A,Zhang-17-Oct-A}.} The energy harvesting efficiency $\eta$ is set to $0.5$. The tolerance level for the convergence of Alg. \ref{alg1} is set to $0.001$ and to calculate the average spectral efficiency, we run {\color{black}$1000$} independent simulations and average the results to get the final figures.

\ifCLASSOPTIONpeerreview
\begin{figure*}[t]
    \centering
    \begin{minipage}[h]{0.48\textwidth}
    \centering
    \includegraphics[width=1.01 \textwidth]{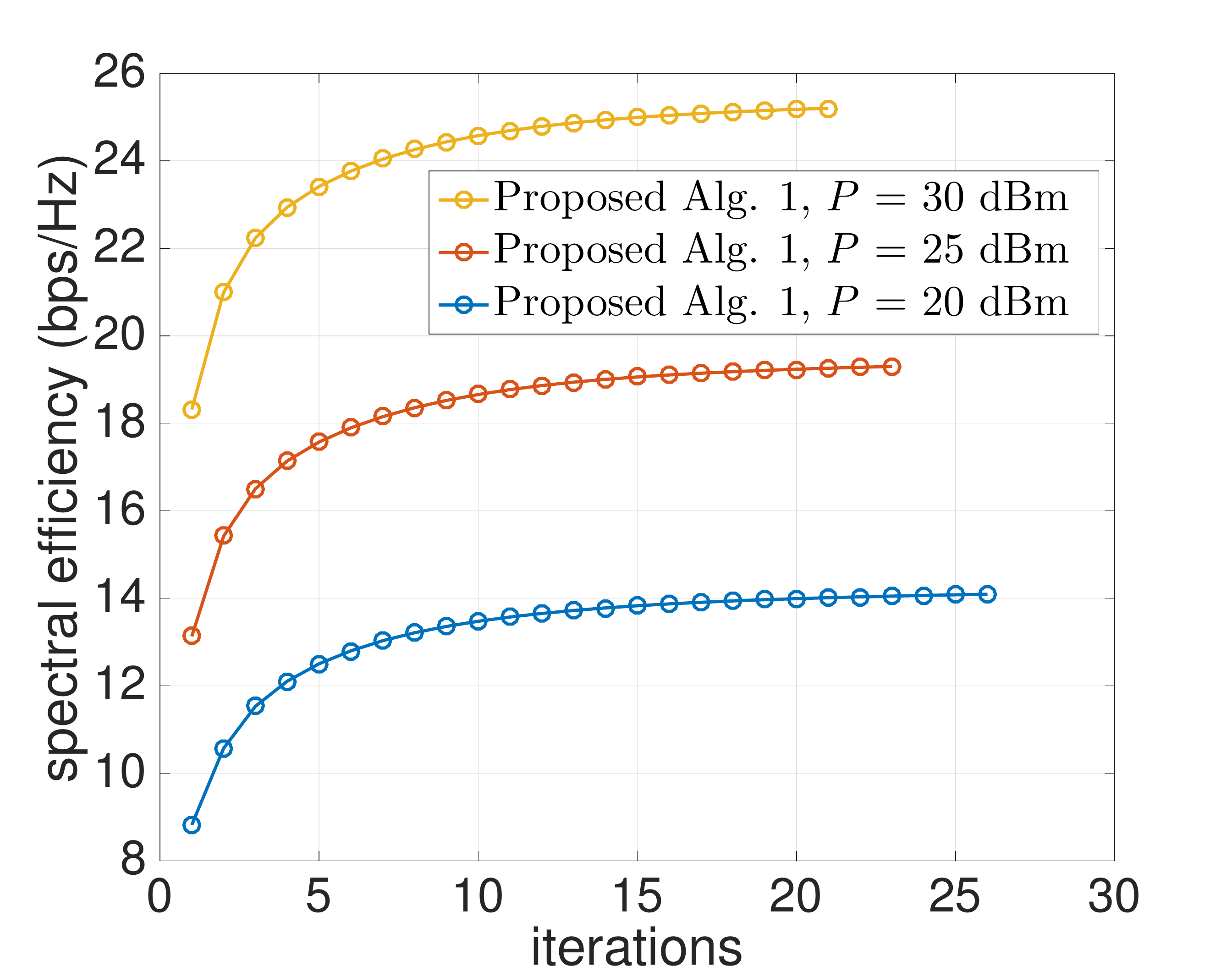}
  \caption{The convergence of Proposed Alg. \ref{alg1}.}
  \label{fig:iter_P}
  \end{minipage}
    \hspace{0.3cm}
    \begin{minipage}[h]{0.48\textwidth}
    \centering
    \includegraphics[width=1.01 \textwidth]{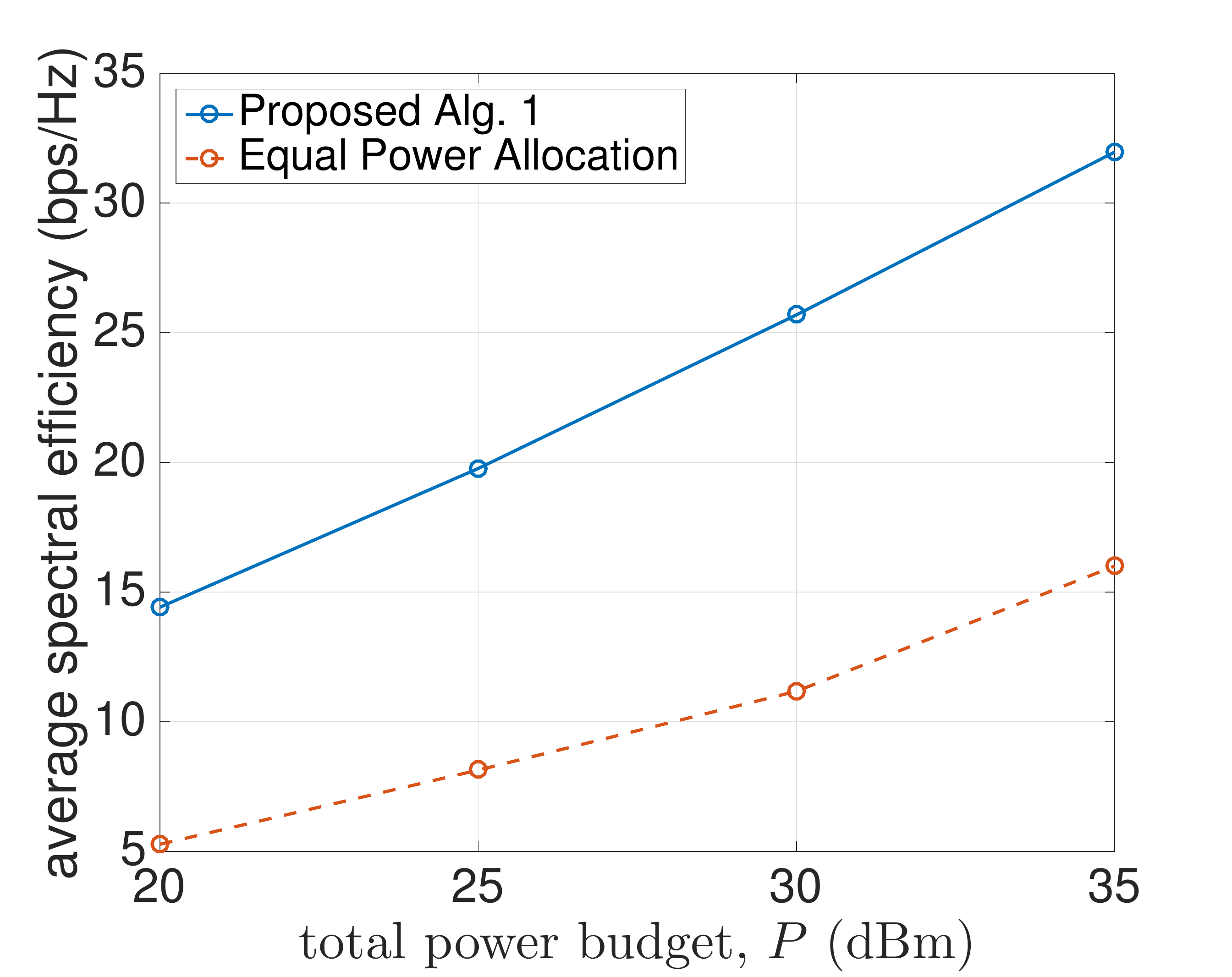}
  \caption{Average spectral efficiency versus the total power budget $P$.}
  \label{fig:r_P}
  \end{minipage}
\end{figure*}
\else
\fi

\ifCLASSOPTIONpeerreview
\else
\begin{figure}[t]
    \centering
    \includegraphics[width=0.48 \textwidth]{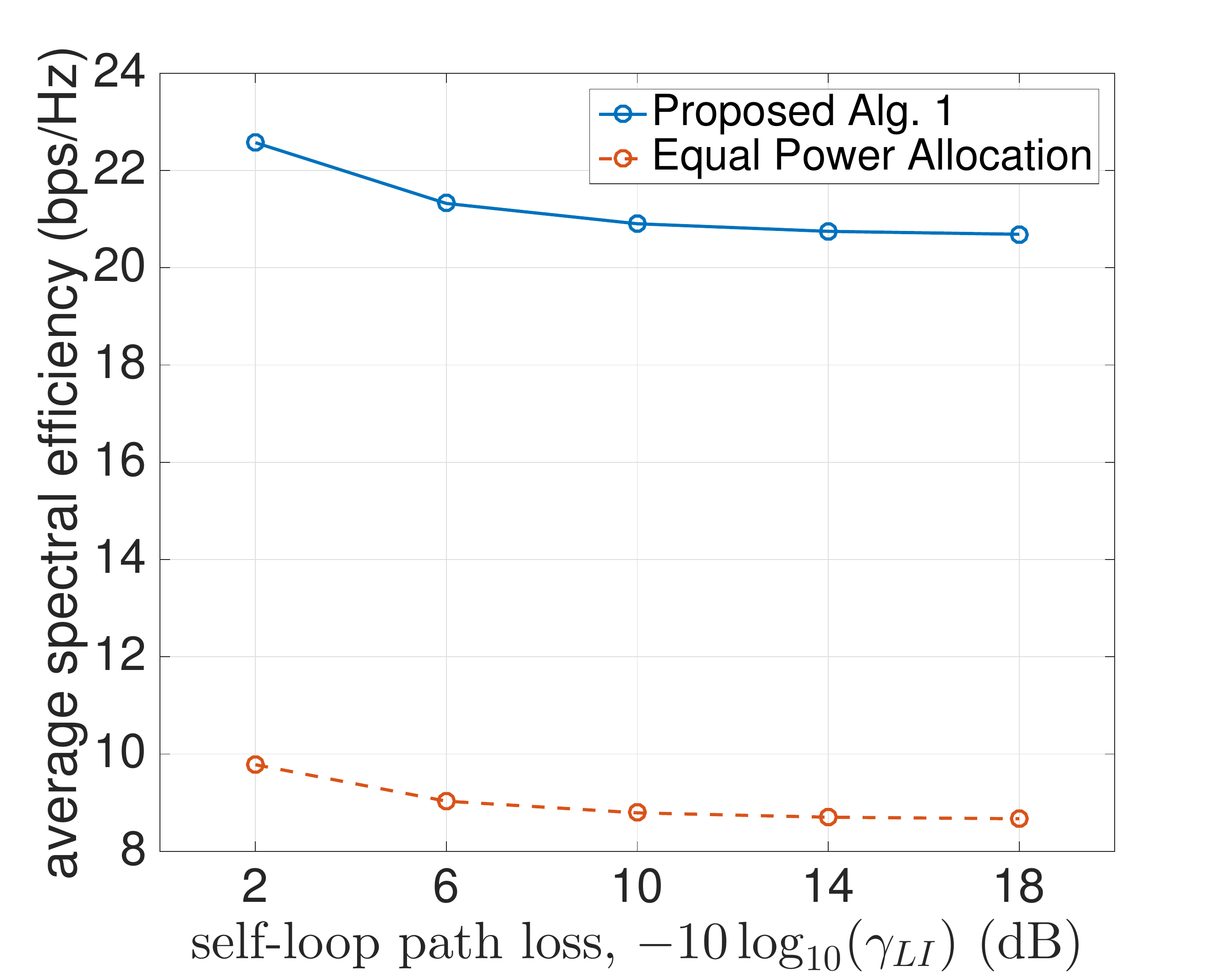}
  \caption{Average spectral efficiency versus the self loop path loss $ \frac{1}{ \gamma_{LI}}$ (evaluated in dB).}
  \label{fig:r_hl}
\end{figure}

\fi

\ifCLASSOPTIONpeerreview
\else
\begin{figure*}[t]
    \centering
    \begin{minipage}[h]{0.48\textwidth}
    \centering
    \includegraphics[width=1.01 \textwidth]{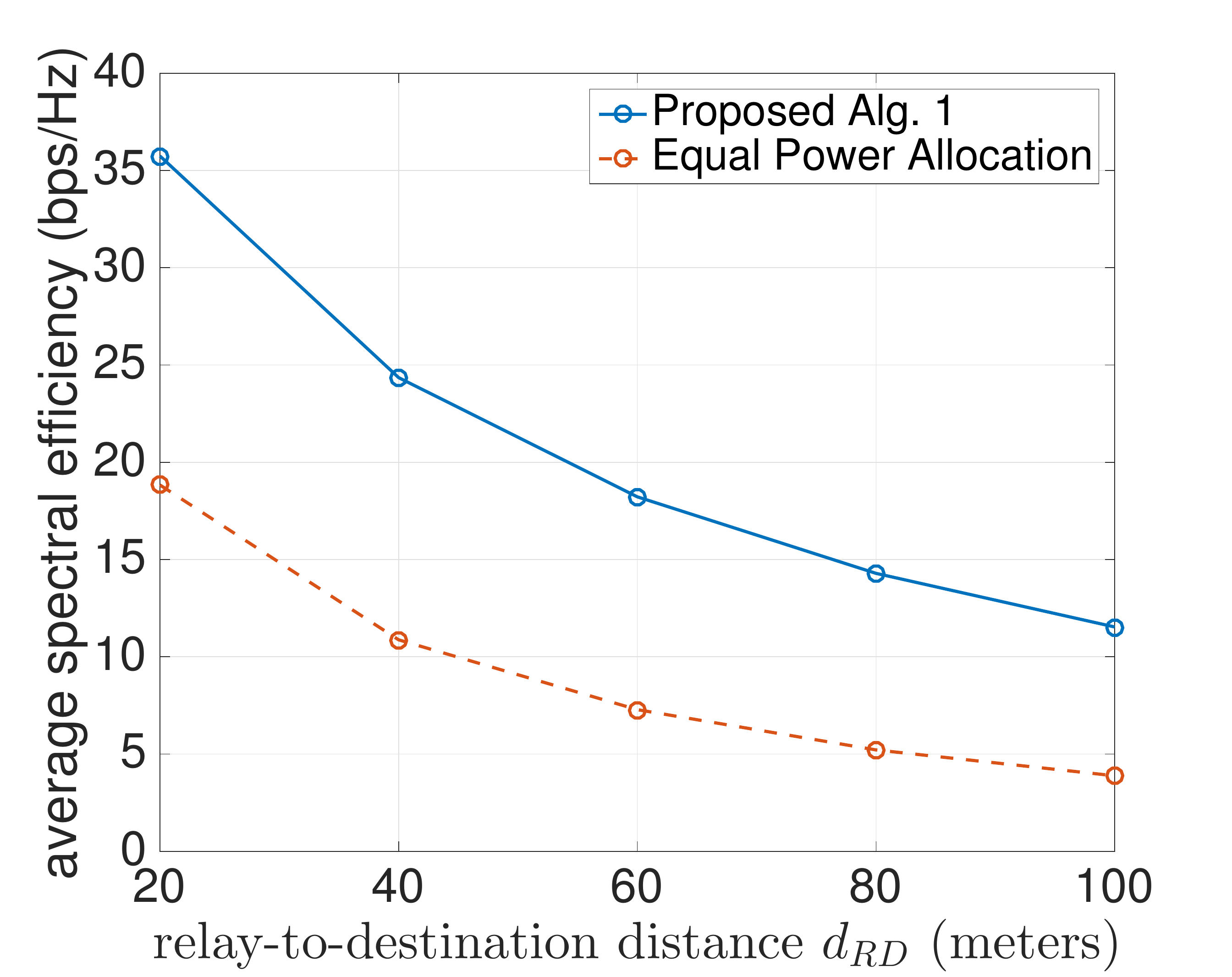}
  \caption{Average spectral efficiency versus the relay-to-destination distance $d_{RD}$.}
  \label{fig:r_d}
  \end{minipage}
    \hspace{0.3cm}
    \begin{minipage}[h]{0.48\textwidth}
    \centering
    \includegraphics[width=1.01 \textwidth]{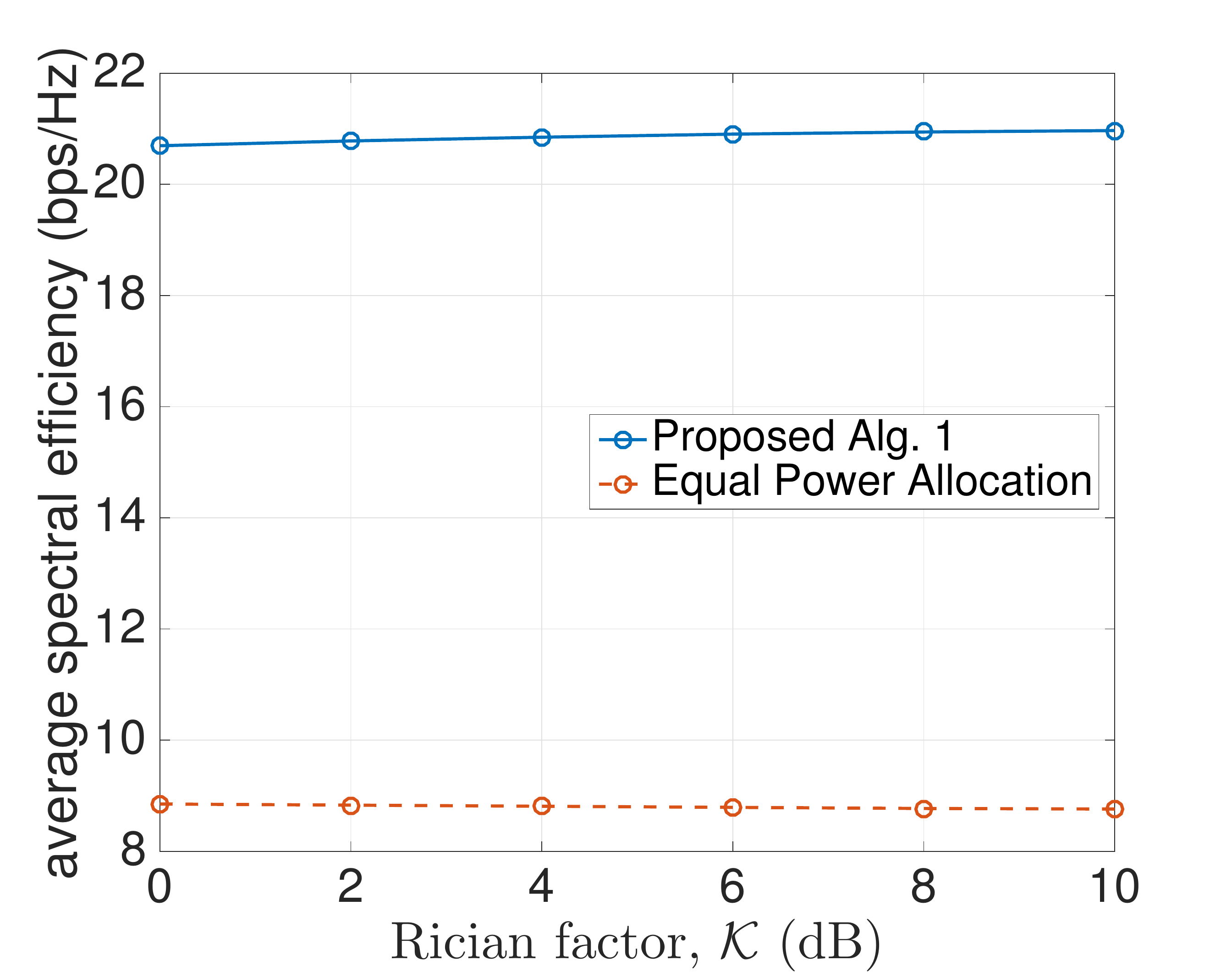}
  \caption{Average spectral efficiency versus the Rician factor $\mathcal{K}$ of the first tap of the time domain source-to-relay channel.}
  \label{fig:r_K}
  \end{minipage}
\end{figure*}
\fi

Fig. \ref{fig:iter_P} shows the convergence of the proposed Alg. 1 for different values of the power budget $P$ for a particular simulation. We can see that the algorithm converges quickly after 20-25 iterations.  On average, Alg. 1 requires $\{26.49,23.67,21.34\}$ iterations before convergence for $P = \{20,25,30\}$ dBm. Fig. \ref{fig:iter_P} shows that the required number of iterations decreases with an increase in the power budget, which means that the Alg. \ref{alg1} quickly figures out the optimal resource allocation in the presence of a relatively large power budget. As expected, increasing the power budget increases the achievable spectral efficiency. This is also shown through Fig. \ref{fig:r_P}, which plots the average spectral efficiency for different values of power $P$. Fig. \ref{fig:r_P} also plots the spectral efficiency results for equal power allocation, which assumes the solution found during initialization of Alg. \ref{alg1}. It can be seen from Fig. \ref{fig:r_P} that the optimization by the proposed Alg. \ref{alg1} achieves significant gain in the spectral efficiency compared to the equal power allocation approach.

\ifCLASSOPTIONpeerreview
\begin{figure}[t]
    \centering
    \includegraphics[width=0.48 \textwidth]{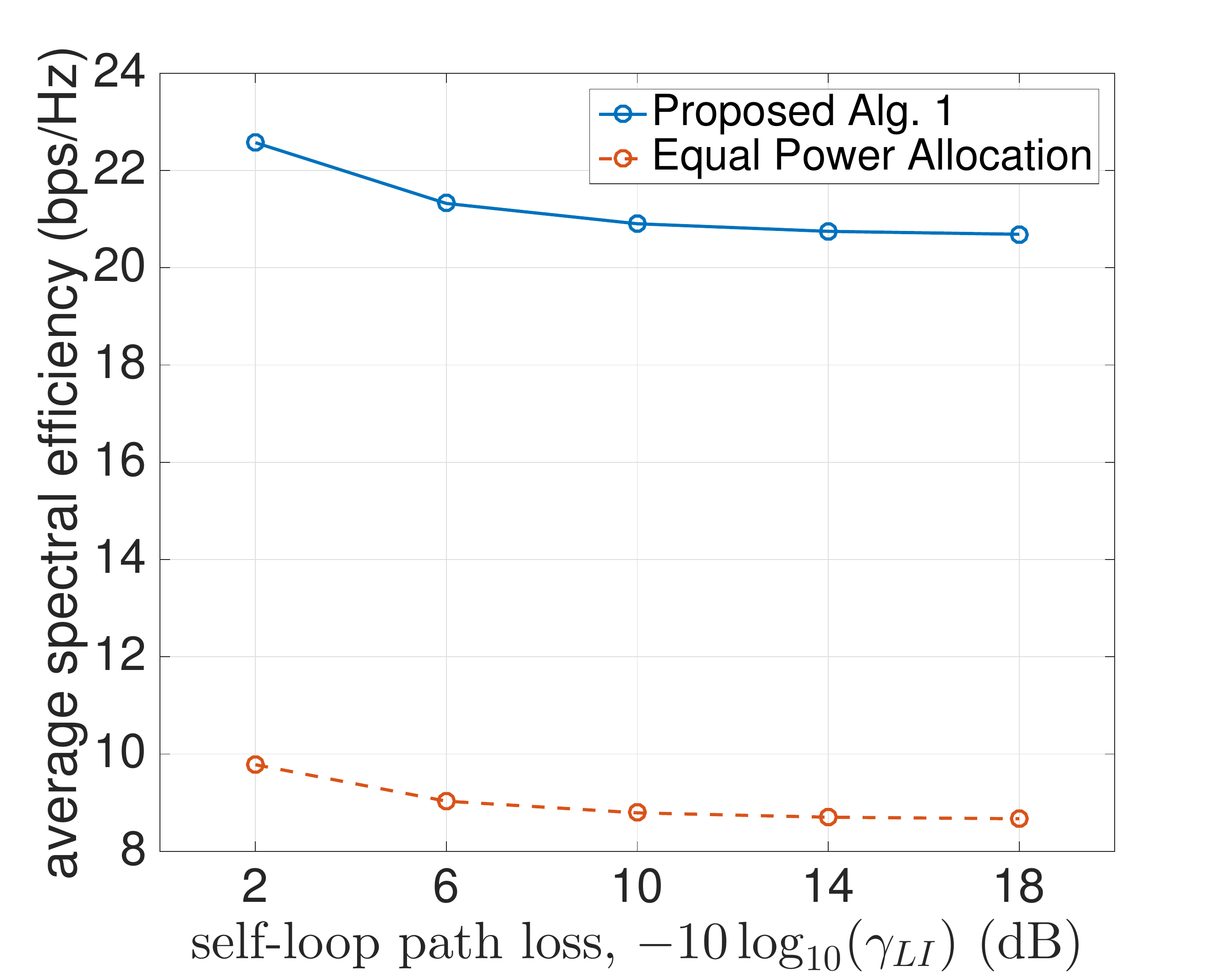}
  \caption{Average spectral efficiency versus the self loop path loss $ \frac{1}{ \gamma_{LI}}$ (evaluated in dB).}
  \label{fig:r_hl}
\end{figure}

\else
\fi

Fig. \ref{fig:r_hl} plots the average spectral efficiency versus the self loop path loss $ \frac{1}{ \gamma_{LI}}$. As expected, the increase in the self-loop path loss decreases the achievable spectral efficiency. This is because increase in the self-loop path loss decreases the amount of harvested energy from the loop self-interference, which decreases the available transmission power from the relay and that would decrease the spectral efficiency at the destination. It is also clear from Fig. \ref{fig:r_hl} that the optimization by the proposed Alg. \ref{alg1} achieves significant gain in the spectral efficiency compared to the equal power allocation approach.

\ifCLASSOPTIONpeerreview
\begin{figure*}[t]
    \centering
    \begin{minipage}[h]{0.48\textwidth}
    \centering
    \includegraphics[width=1.01 \textwidth]{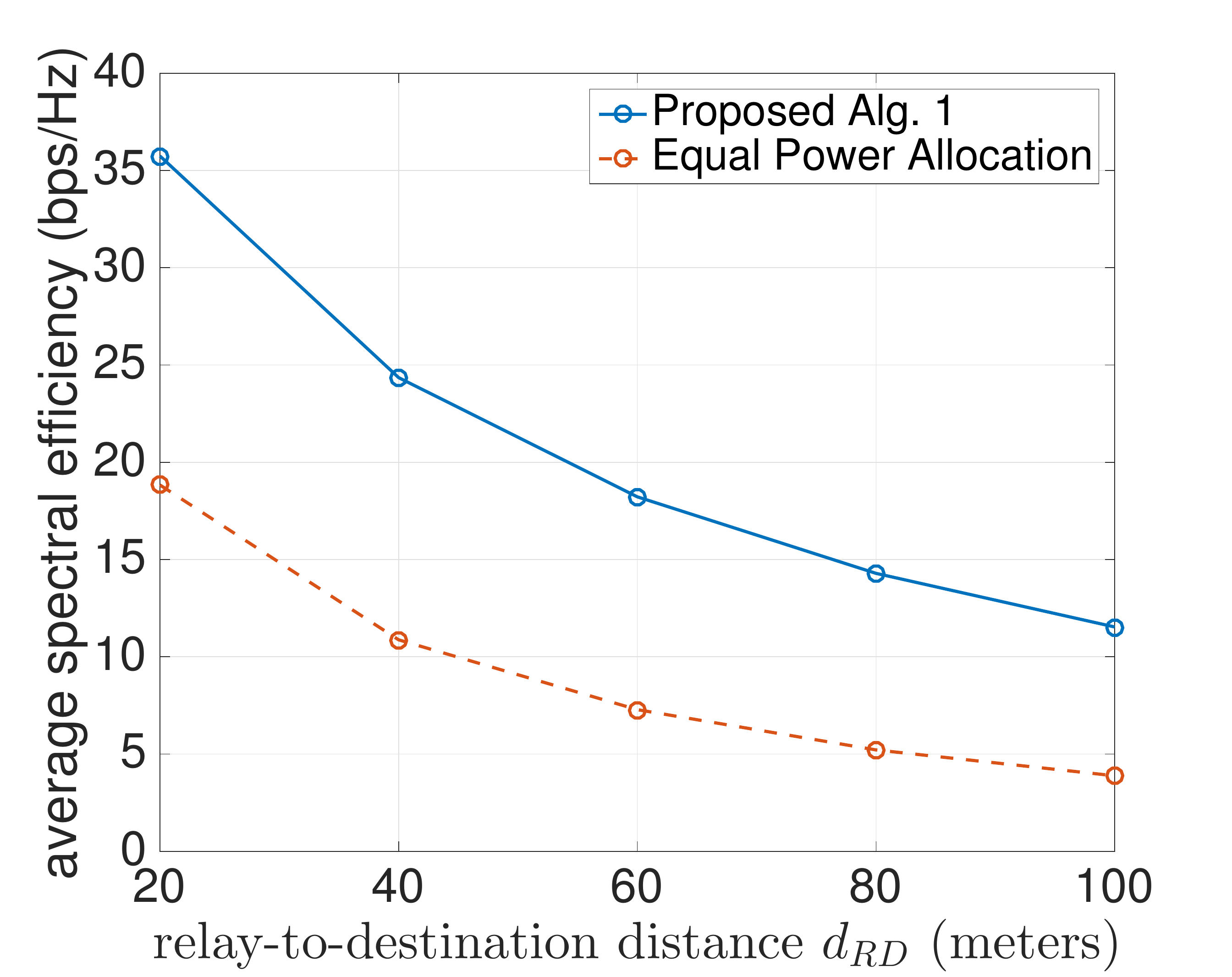}
  \caption{Average spectral efficiency versus the relay-to-destination distance $d_{RD}$.}
  \label{fig:r_d}
  \end{minipage}
    \hspace{0.3cm}
    \begin{minipage}[h]{0.48\textwidth}
    \centering
    \includegraphics[width=1.01 \textwidth]{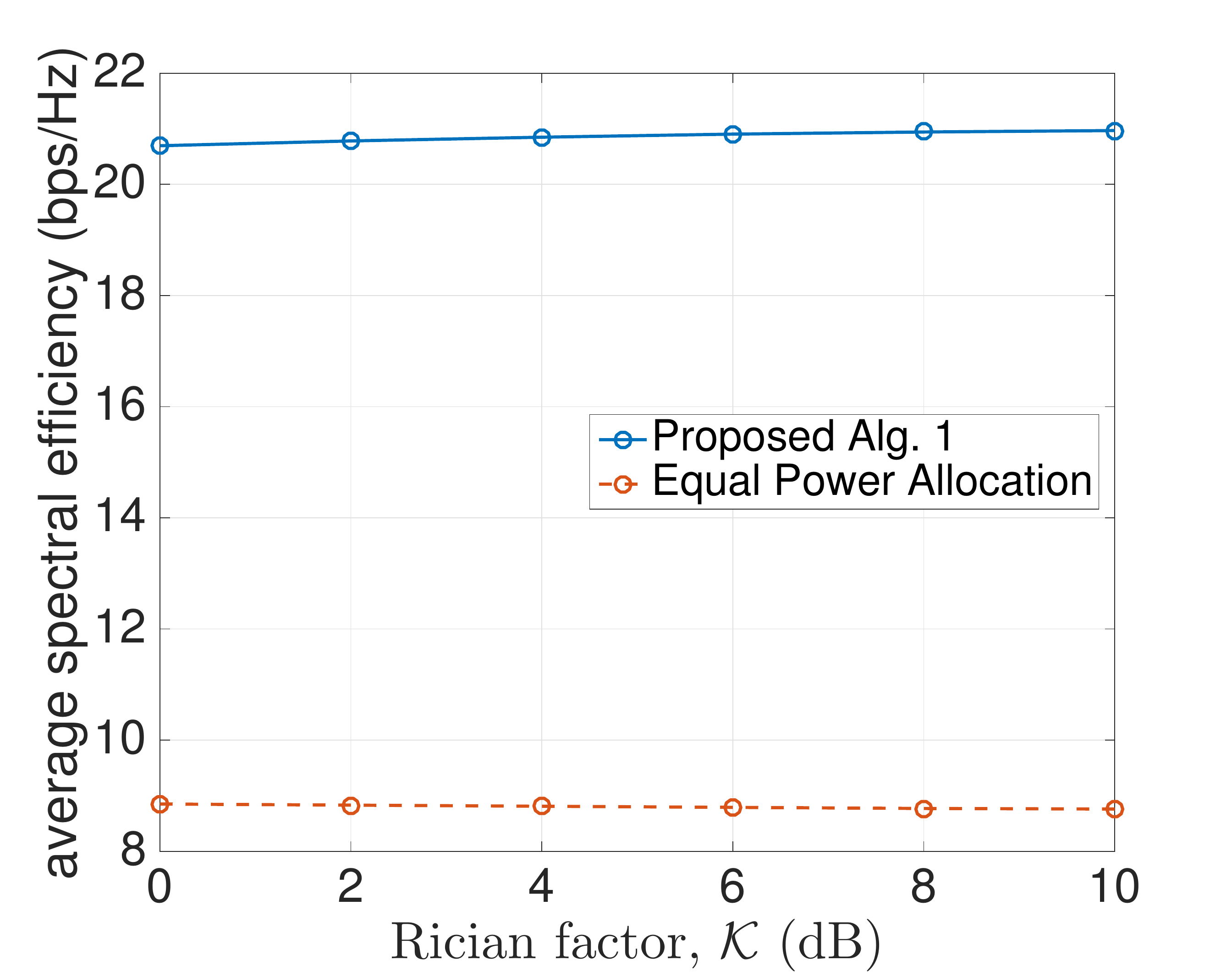}
  \caption{Average spectral efficiency versus the Rician factor $\mathcal{K}$ of the first tap of the time domain source-to-relay channel.}
  \label{fig:r_K}
  \end{minipage}
\end{figure*}
\else
\fi

Fig. \ref{fig:r_d} plots the average spectral efficiency versus the relay-to-destination distance $d_{RD}$. As expected, the increase in the relay-to-destination distance decreases the achievable spectral efficiency due to increase in the path-loss. Fig. \ref{fig:r_K} plots the average spectral efficiency versus the Rician factor $\mathcal{K}$ of the first tap of the time domain source-to-relay channel. {\color{black}Fig. \ref{fig:r_K} shows that increase in the Rician factor results in a very slight increase in the achievable spectral efficiency. This is because only the first channel tap of the time-domain source-to-relay channel is Rician distributed and the rest of the channel taps of source-to-relay link and all the channel taps of relay-to-destination link are Rayleigh distributed.} It is also clear from Figs. \ref{fig:r_d}-\ref{fig:r_K} that the optimization by the proposed Alg. \ref{alg1} achieves significant gain in the spectral efficiency compared to the equal power allocation approach. {\color{black} This significant gain is due to the frequency selective nature of the channel, where equal power allocation suffers a lot compared to the optimal resource allocation.}

\ifCLASSOPTIONpeerreview
\begin{table}[ht]
\caption{Computational time of Proposed. Alg. 1 in seconds.} 
\centering 
\begin{tabular}{| c ||c |c |c |c |} 
\hline
Solving Methodology & $K = 64$ & $K = 256$ & $K = 1024$ & $K = 4096$ \\ [0.5ex] 
\hline   
Proposed. Alg. 1 & 0.010 sec. & 0.017 sec. & 0.046 sec. & 0.1602 sec.  \\ [0.5ex]
\hline 
\end{tabular}
\label{table:comp_time} 
\end{table}
\else
\fi

\ifCLASSOPTIONpeerreview
\else
\begin{table*}[ht]
\caption{Computational time of Proposed. Alg. 1 in seconds.} 
\centering 
\begin{tabular}{| c ||c |c |c |c |} 
\hline
Solving Methodology & $K = 64$ & $K = 256$ & $K = 1024$ & $K = 4096$ \\ [0.5ex] 
\hline   
Proposed. Alg. 1 & 0.010 sec. & 0.017 sec. & 0.046 sec. & 0.1602 sec.  \\ [0.5ex]
\hline 
\end{tabular}
\label{table:comp_time} 
\end{table*}
\fi

\begin{figure}[t]
    \centering
    \includegraphics[width=0.48 \textwidth]{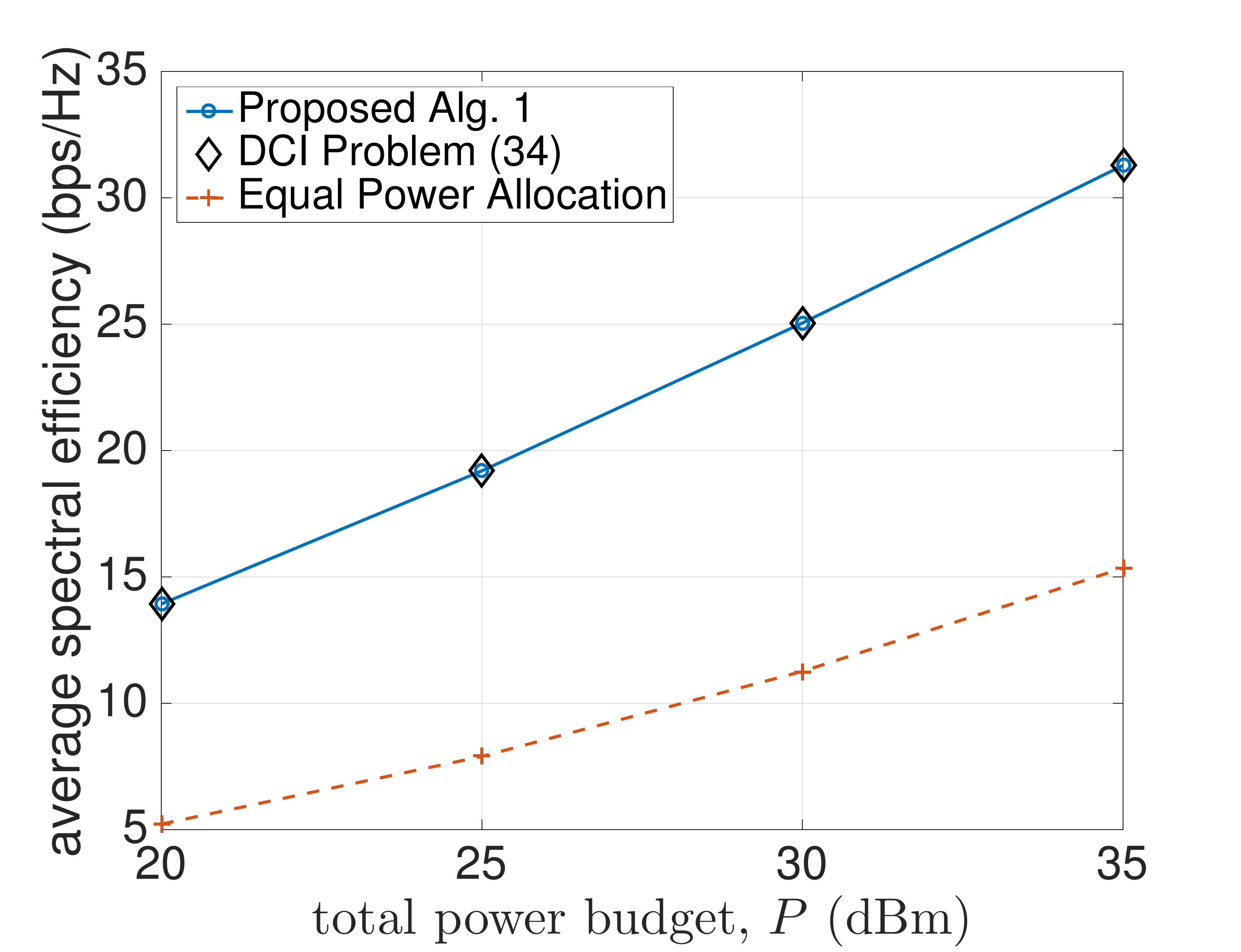}
  \caption{Comparison of average spectral efficiency versus the total power budget $P$ for different algorithms.}
  \vspace*{-.2cm}
  \label{fig:r_P_comp}
\end{figure}

{\color{black} { \it Computational Complexity and Performance Comparison:} }

The closed-form solution proposed in Alg. \ref{alg1} is computationally very efficient. The computational time of the proposed Alg. 1 is shown in Table \ref{table:comp_time} for different problem conventionality. The computation time is calculated on a $2.7$ GHz Intel core-i$5$ machine with $8$ GB RAM. Table \ref{table:comp_time} shows that the closed-form solution proposed in Alg. \ref{alg1} quickly finds the optimal solution (in less than $\frac{1}{10}$ second) and there is very slight increase in the computational time as the scale of the problem grows (as the number of subcarriers $K$ increases). Note that this closed-form solution proposed in Alg. \ref{alg1} is quite efficient compared to directly solving of Problem \eqref{p2} {\color{black}(or DCI Problem \eqref{dc_prob})} using CVX, as the latter approaches requires around $32$ minutes {\color{black}(or $26$ minutes)} even for a very small-scale problem ($K = 64$).

{\color{black} In Fig. \ref{fig:r_P_comp}, we compare the average spectral efficiency of the proposed algorithm with that obtained by solving the DCI Problem \eqref{dc_prob}. To ensure reasonable simulation time for solving the DCI Problem \eqref{dc_prob} using CVX, we consider a small-scale problem, $K = 16$. In order to simulate small-scale system with $K = 16$ sub-carriers or even for $K = 64$ sub-carriers, we generate the channel with an exponential power delay profile having root-mean-square delay spread of $\sigma_\text{RMS} = T_s$, which results in $L=4$-tap channel. It can be seen from Fig. \ref{fig:r_P_comp} that the proposed Alg. \ref{alg1} achieves similar spectral efficiency performance as that obtained by the DCI Problem \eqref{dc_prob}. Note that we also achieve similar spectral efficiency performance by solving the Problem \eqref{p2} because it has similar structure compared to the DCI Problem \eqref{dc_prob} as explained in Remark 3. In the following, we elaborate through general computational complexity analysis that by how-much factor, our proposed solution in Alg. \ref{alg1} is efficient compared to the DCI Problem \eqref{dc_prob}.

In general, the computational complexity of the proposed algorithm is $\mathcal{O} \left( i_\text{A1} (18 K N + 39) \right)$ and the computational complexity of the DCI Problem \eqref{dc_prob} is  $\mathcal{O} \left( i_\text{DC}  ( (3KN)^3(3KN+2) ) \right)$ \cite{Peaucelle-02-A}, where $i_\text{A1}$ is the average number of iterations required for the convergence of the proposed Alg. \ref{alg1} and $i_\text{DC}$) is the average number of iterations required for the convergence of the DCI Problem \eqref{dc_prob}. Our simulations show quite similar values of $i_\text{A1}$ and $i_\text{DC}$, e.g., for the small-scale problem ($K = 16$) considered in Fig. \ref{fig:r_P_comp}, $i_\text{A1} = 25.8$ and  $i_\text{DC} = 25.7$ for $P = 20$ dBm. However, it is noteworthy that that our proposed closed-form solution in Alg. \ref{alg1} is about $(KN)^3$ times computationally faster than solving the DCI Problem \eqref{dc_prob}. }

\section{Conclusions}
This paper has considered a MIMO-OFDM system with FD relaying, in which the cooperative relay forwards the source information to the destination and in meanwhile, replenishes or harvests energy not only from wireless signals from the source but also through energy recycling from its own transmission. The high values of residual self-interference are helpful for recharging of the energy-constrained relay node. The objective is to maximize the spectral efficiency of such a system and come up with the optimal power allocation over each sub-carrier and each transmit antenna. Such power allocation is very challenging in the presence of a large number of sub-carriers due to large-scale optimization. Thus, we have proposed a new path-following algorithm, which is practical for large-scale optimization. Our approach is computationally very efficient and practical since the proposed solution requires only a few \textit{closed-form} calculations. Our numerical results with a practical simulation setup show promising results by achieving high spectral efficiency. Moreover, the achieved spectral efficiency is quite high compared to the simple ``equal power allocation" approach.

{\color{black} It could be the subject of future research to investigate that how the proposed Alg. \ref{alg1} can be used for the application of 5G new radio, which is based on modified OFDM waveform. Moreover, the considered problem in this work can be solved with practical channel estimation in the presence of channel uncertainties.}

\section*{Appendix: basic inequalities}
As the function $\gamma(x)=\ln(1+x)$ is concave in the domain $\mbox{dom}(\gamma)= \{x>0\}$, it is true
that \cite{Tuybook}
\begin{eqnarray}\label{ap2}
\ln(1+x) &\leq& \ds \gamma(\bar{x})+\frac{\partial \gamma(\bar{x})}{\partial x}(x-\bar{x})\nonumber\\
&=&\ln(1+\bar{x})+\ds\frac{x-\bar{x}}{1+\bar{x}}\quad\forall\ x>0, \bar{x}>0.
\end{eqnarray}
On the other hand, function $\beta(y)\triangleq \ln(1+e^y)$ is convex in the domain
$\mbox{dom}(\beta)=\{y>0\}$ because $\partial^2 f(y)/\partial y^2=e^y/(1+e^y)^2>0$ $\forall\ y\in\mbox{dom}(\beta)$.
Therefore \cite{Tuybook}
\begin{eqnarray}\label{ap1a}
\ln(1+e^y)&\geq& \beta(\bar{y})+\ds\frac{\partial \beta(\bar{y})}{\partial y}(y-\bar{y})\nonumber\\
&=&\ln(1+e^{\bar{y}})+\ds\frac{e^{\bar{y}}}{1+e^{\bar{y}}}(y-\bar{y}).
\end{eqnarray}
Substituting $x=e^y$ and $\bar{x}=e^{\bar{y}}$ into (\ref{ap1a}) yields the following inequality
\begin{equation}\label{ap1}
\ds \ln(1+x)\geq  \ln(1+\bar{x})+\frac{\bar{x}}{1+\bar{x}}\left(\ln x-\ln\bar{x} \right)\quad\forall\
x>0, \bar{x}>0.
 \end{equation}

 \ifCLASSOPTIONpeerreview
\else
\balance
\fi


\end{document}